\documentclass{aa}  

\usepackage{graphicx}
\usepackage[varg]{txfonts}
\usepackage{hyperref}

\usepackage{natbib}
\bibpunct{(}{)}{;}{a}{}{,}

\begin{document}

	\title{A simple two-component description of energy equipartition and mass segregation for anisotropic globular clusters}
	
	\author{S. Torniamenti\inst{1}
		\and
		G. Bertin\inst{1}
		\and
		P. Bianchini\inst{2}
	}
	
	\institute{
		Universit\`{a} degli Studi di Milano, Dipartimento di Fisica, Via Celoria 16, 20133 Milano, Italy 
		\and
		Observatoire Astronomique de Strasbourg, 11 Rue de l'Universit\'{e}, 67000 Strasbourg, France 
	}
	
	\date{Received <date> /
		Accepted <date>}

	\abstract
	{In weakly-collisional stellar systems such as some globular clusters, partial energy equipartition and mass segregation are expected to develop as a result of the cumulative effect of stellar encounters, even in systems initially characterized by star-mass independent density and energy distributions. In parallel, numerical simulations have demonstrated that radially-biased pressure anisotropy slowly builds up in realistic models of globular clusters   from initial isotropic conditions, leading to anisotropy profiles that, to some extent, mimic those resulting from incomplete violent relaxation known to be relevant to elliptical       galaxies. In this paper, we consider a set of realistic simulations realized by means of Monte Carlo methods and analyze them by means of self-consistent, two-component models.  For this purpose, we refer to an underlying distribution function originally conceived to describe elliptical galaxies, which has recently been truncated and adapted to the context of globular clusters. The two components are supposed to represent light stars (combining all main sequence stars) and heavy stars (giants, dark remnants, and binaries). We show that this conceptually simple family of two-component truncated models provides a reasonable description of simulated density, velocity dispersion, and anisotropy profiles, especially for the most relaxed systems, with the ability to quantitatively express the attained levels of energy equipartition and mass segregation.  In contrast, two-component isotropic models based on the King distribution function do not offer a comparably satisfactory representation of the simulated globular clusters. With this work, we provide a new reliable diagnostic tool applicable to nonrotating globular clusters that are characterized by significant gradients in the local value of the mass-to-light ratio, beyond the commonly used one-component dynamical models. In particular, these models are supposed to be an optimal tool for the clusters that underfill the volume associated with the boundary surface determined by the tidal interaction with the host galaxy.}
	
	%
	
	\titlerunning{A simple two-component description for anisotropic globular clusters}
	\maketitle
	
	\section{Introduction}
	
	The underlying strategy for the construction of several physically-based dynamical models of stellar systems may be summarized in the following steps. A picture of formation and evolution for the stellar systems under consideration is adopted. As for the dynamical mechanisms that are involved, and to the end-products of the formation scenario, the picture is further explored by means of dedicated simulations.  The simulations provide clues on the phase-space structure of the stellar systems under investigation, generally well beyond the reach of direct observations. These clues are used to identify (possibly simple) candidate distribution functions to incorporate the desired dynamical features, and to construct self-consistent solutions from the Vlasov-Poisson system of equations. A comparison with the observed stellar systems is performed by converting the properties of the selected dynamical models into surface brightness profiles and velocity dispersion profiles, under the assumption that the stellar populations are characterized by a constant mass-to-light ratio within a given system. If the comparison with the observations is reasonably good, the simple models can be used to fit the data and to infer, for a given system, many structural properties that observations are unable to determine directly. The models are then made more complex if required by some astrophysical issues.
	
	For elliptical galaxies, families of models developed according to the above description are based on the assumption that those stellar systems are the result of incomplete violent relaxation (\citealp{1967MNRAS.136..101L}, \citealp{1982MNRAS.201..939V}). The properties of the stellar populations of elliptical galaxies, and the mechanism of  violent relaxation, naturally encourage the use of the assumption of a constant mass-to-light ratio for the visible matter; eventually, the models were extended to include the presence of a dark halo (e.g., see \citealp{1993RPPh...56..493B}).
	
	For globular clusters, the most widely used models, the King models (\citealp{1966AJ.....71...64K}) were developed under the guiding picture that for many clusters, given their long age, two-body relaxation processes have had sufficient time to act and play an important role in determining the dynamical properties of the systems that we observe today. King models are thought to describe round, non-rotating stellar systems made of a single stellar population, for which internal two-body relaxation has had time to bring the system close to a Maxwellian, isotropic distribution function. A truncation is considered to take into account the presence of tidal effects. The success of one-component King models was largely due to their simplicity (the family of models is characterized by only one dimensionless parameter, the concentration parameter $c$) and to their ability to fit, under the assumption of a constant mass-to-light ratio, the photometric profiles of most clusters in our Galaxy over large radial/magnitude ranges (e.g., see \citealp{1994AJ....108.1292D}, \citealp{1996yCat.7195....0H}). In the absence of accurate and radially-extended kinematical data, King models have been used to infer several structural properties of globular clusters beyond the reach of observations. Unfortunately, the processes of relaxation that set the foundation of the King models include qualitative elements that were immediately recognized to be discordant, at least in principle, with a simple description in terms of one-component models. 
	
	One of the effects of two-body interactions is to lead systems made of stars with different masses toward a state of energy equipartition. That is to say, we would expect collisions to enforce a condition in which the velocity dispersion $\sigma$ of stars of mass $m$ should scale as $\sigma \sim m^{-1/2}$.  In self-gravitating    systems, the establishment of this process is very complicated, because of their inhomogeneous nature and their self-consistent dynamics. In turn, as a consequence of two-body relaxation processes, more massive stars are expected to be characterized by a more concentrated density distribution, a phenomenon usually referred to as mass segregation. This was soon realized to pose a contradiction to the simple use of one-component models in the interpretation of the observations and efforts were made to extend the King models to a more realistic multi-component version (\citealp{1976ApJ...206..128D}; see also \citealp{1981AJ.....86..318M}). On the dynamical side, especially because gravity is a natural source of inhomogeneities, it was also found that interesting dynamical mechanisms might be induced by the natural trend toward equipartition and mass segregation. In particular, arguments were provided in favor of the existence of an instability related to mass segregation (\citealp{1969ApJ...158L.139S}), distinct from the gravothermal catastrophe (\citealp{1968MNRAS.138..495L}). With the help of a very simple two-component model, made of light and heavy stars, Spitzer suggested that a condition of global energy equipartition cannot be fulfilled if the total mass of the heavy stars exceeds a certain fraction of the total mass of the cluster. The Spitzer criterion (often interpreted as a criterion for instability) was later extended by \cite{1978ApJ...223..986V} to cover the case of a continuous spectrum of masses. In general, as shown by the discussion that followed the study by Spitzer (e.g., see \citealp{1981AJ.....86..318M}), energy equipartition and, in particular, the distinction between local and global equipartition, are subtle concepts that require clarification.
	
	The advent of improved numerical simulations and of increasingly better observations have confirmed that the general stellar dynamical modeling procedure for globular clusters requires a substantial upgrade.  Numerical experiments confirmed that globular clusters can attain a condition of only partial energy equipartition even in their central, most relaxed, regions (\citealp{2013MNRAS.435.3272T}, \citealp{2016MNRAS.458.3644B}, see also \citealp{1981AJ.....86..318M}, \citealp{2006MNRAS.366..227M}). A curious and rather unexpected (but see \citealp{1971Ap&SS..13..284H}) phenomenon demonstrated by numerical simulations is the fact that, as a result of the slow action of two-body relaxation, radially-biased pressure anisotropy slowly builds up in realistic models of globular clusters also from initially isotropic conditions (\citealp{2016MNRAS.461..402T}, \citealp{2016MNRAS.462..696Z}, \citealp{2017MNRAS.471.1181B}), leading to anisotropy profiles that, to some extent, mimic those resulting from incomplete violent relaxation, known to be relevant to elliptical galaxies.
	
	On the observational side, with the measurement of proper motions by the Hubble Space Telescope (see the HSTPROMO data sets for 22 globular clusters, \citealp{2014ApJ...797..115B}, \citealp{2015ApJ...803...29W}; interesting related results are expected to come from the \textit{Gaia} mission, \citealp{2018A&A...616A..12G}), it has been confirmed that a state of only partial energy equipartition is indeed established in globular clusters (e.g., see \citealp{2013MNRAS.435.3272T}, \citealp{2018ApJ...861...99L}, \citealp{2019ApJ...873..109L}). In addition, evidence for a certain degree of mass segregation has been collected for several globular clusters (e.g., see \citealp{2010ApJ...710.1063V}, \citealp{2013AJ....145..103D},   \citealp{2013ApJ...778...57G}, \citealp{2014ApJ...797..115B}, \citealp{2017MNRAS.464.1977W}). In relation to the velocity space, one major surprise has been the finding of significant (differential) rotation in some globular clusters (\citealp{2003AJ....126..772A},   \citealp{2012A&A...538A..18B}, \citealp{2013ApJ...772...67B},  \citealp{2014A&A...567A..69K},  \citealp{2015A&A...573A.115L}, \citealp{2017ApJ...844..167B}, \citealp{2017MNRAS.465.3515C}, \citealp{2018MNRAS.481.2125B}, \citealp{2018MNRAS.473.5591K},   \citealp{2019MNRAS.485.1460S}), which brings us well beyond the goals of the present paper. Besides the issue of rotation, the new kinematical data confirm the presence of pressure anisotropy (\citealp{2017ApJ...844..167B}, \citealp{2015ApJ...803...29W}, \citealp{2019arXiv190311070J}). 
	
	In this general context, we wish to look for an improved multi-component upgrade of the stellar-dynamical modeling of globular clusters by taking into account the combined problems of energy equipartition, mass segregation, and pressure anisotropy. Beyond the traditional King models, a variety of models have been constructed and applied to the study of globular clusters. Among the interesting models characterized by pressure anisotropy, we can mention the Michie-King models (\citealp{1963MNRAS.125..127M}) and the LIMEPY models (\citealp{2015MNRAS.454..576G}). Multi-component versions of these models have been tested with some success both as a description of data and of the results of N-body simulations (\citealp{1979AJ.....84..752G}, \citealp{2015MNRAS.451.2185S}, \citealp{2017MNRAS.470.2736P}). The general goal of these studies appears to be  to propose realistic models by using the freedom offered by the presence of some additional parameters (see the pioneering work by \citealp{1976ApJ...206..128D}). If we give priority to simplicity, which would be a welcome factor both in view of applications to the study of dynamical mechanisms (such as the so-called Spitzer instability), and of the development of a diagnostic tool for comparison with the observations, we can try to start from a two-component description (based on light and heavy stars). Given the experience with the modeling of elliptical galaxies, a first attempt in this direction was recently made by \cite{2016A&A...590A..16D}. The family of models introduced in that paper allows us to take into account the presence of partial energy equipartition and mass segregation with simple analytic tools. As shown in that article, the proposed models possess some appealing and promising features. Yet, we are aware that lumping together into two single components of light and heavy stars, each of which corresponds to populations made of individual objects characterized by a large spread of masses and mass-to-light ratios (see Sect. \ref{sec_2c_definition}), might be a bold step, and subject to unacceptable limitations. Therefore, before declaring that the proposed conceptually simple models may offer a reliable tool to study some dynamical mechanisms or to interpret the observations (with the required discussion of the size and role of mass-to-light gradients; see \citealp{2017MNRAS.469.4359B},  Sect. 4.1 of \citealp{2016A&A...590A..16D}, and \citealp{2019MNRAS.483.1400H} for a three-component description), we need to test whether the simple two-component models do indeed have the power to describe the structural properties of the extremely complex globular clusters, as is nowadays well-described by state-of-the-art simulations.
	
	The main objective of this paper is to test the quality of the family of truncated two-component $f^{(\nu)}$ models introduced by \cite{2016A&A...590A..16D}  by checking their properties against those measured in a set of eight snapshots taken from realistic simulations of globular clusters. The simulations under consideration were performed using Monte Carlo methods (\citealp{2010MNRAS.407.1946D}) and have been extensively used to characterize the internal properties of simulated globular clusters with different levels of internal relaxation (\citealp{2016MNRAS.458.3644B}, \citealp{2017MNRAS.469.4359B}). In order to be comprehensive, we carried out a parallel test based on one-component King models and on two-component King  and Michie-King models to test to what extent the adopted family of models does indeed represent an improvement with respect to other natural options.         If our models turn out to perform in a satisfactory way, they will be applicable as a tool to clarify the complex onset of mass segregation and energy equipartition in globular clusters, and possibly to infer the dynamical properties of the subcomponents that cannot be observed. A second important objective of this paper is to compare the anisotropy profiles generated by collisions to those produced by violent relaxation, in order to find possible analogies between the degrees of anisotropies generated by these two different mechanisms.
	
	This paper is organized as follows. In Section 2, we briefly recall the properties of the models used in our investigation. In Section 3, we summarize the main characteristics of the set of Monte Carlo cluster simulations and define the set of simulated states (that we referred to earlier as snapshots) considered in this paper. In Section 4, we test the quality of our two-component models on these states. A discussion and our conclusions are featured in Section 5.
	
	\begin{table*}[ht]
		\caption{Initial conditions of simulations. In the first column, the original label of the simulations from \cite{2010MNRAS.407.1946D} is given in parentheses. The other columns list the binary fraction $f_{binary}$, the ratio of the truncation to the half-mass radius $r_t/r_M$, the total number of particles N (defined as the number of single stars plus the number of binary systems), the total mass $M_{tot}$, the truncation radius $r_t$, and the half-mass relaxation time defined as in \cite{2010MNRAS.407.1946D}.}             
		\label{tab_bianchini_init}      
		\centering                          
		\resizebox{\textwidth}{!}{
			\begin{tabular}{l c c c c c c c}        
				\hline   \hline              
				&$f_{binary} \, (\%)$ & $r_t/r_M$ & N &$M_{tot} \, [M_{\odot}]$ & $r_t \, [pc]$  & $t_{rh} [Gyr]$ \\    
				\hline                        
				Sim 1 (10low75) & 10  & 75 & $5 \times 10^5$ & $3.62 \times 10^5$ & 150 & $0.53$   \\
				Sim 2 (50low75) & 50  & 75  & $5 \times 10^5$ & $5.07 \times 10^5$ & 150 & $0.44 $   \\
				Sim 3 (10low37) & 10  & 37 & $5 \times 10^5$ & $ 3.62 \times 10^5$ & 150 & $1.51$   \\
				Sim 4 (50low37) & 50 & 37 & $5 \times 10^5$ & $ 5.07  \times 10^5$ & 150 & $1.28$    \\ 
				Sim 5 (10low180) & 10  & 180 & $5 \times 10^5$ & $ 3.62 \times 10^5$ & 150 & $0.14$   \\
				Sim 6 (50low180) & 50 & 180 & $5 \times 10^5$  & $ 5.07  \times 10^5$ & 150 & $0.12$   \\
				Sim 7 (10low75-2M) & 10  & 75 & $20 \times 10^5$ & $7.26 \times 10^5$ & 150 &   \\
				\hline                                   
			\end{tabular}}
		\end{table*}
		
		\section{Two-component models} \label{sec_2cmodels}
		In the spirit of a number of previous papers, such as the one by \cite{1969ApJ...158L.139S}, we tried to model an extremely complex stellar system with a full spectrum of masses by means of simple, idealized two-component models. At variance with the study of large collisionless stellar systems, at least two components were required here, because we are interested in investigating collisional effects that depend on the masses of the interacting stars. With the help of a number of realistic simulations, to be described in the next section, we wish to test in detail how well such a simple, idealized (but physically justified) description is able to capture the structural properties of a realistic model of globular cluster. In the context of the dynamical modeling of globular clusters, a direct comparison between models and simulated states was recently made by \cite{2015MNRAS.451.2185S}, who mostly focused on the issue of the bias in mass estimates obtained from application of four-component Michie-King (\citealp{1963MNRAS.125..127M}) models, and by \citealp{2019MNRAS.483.1400H}, who aimed to compare the performance of a vast variety of mass-modeling techniques on the specific case of an N-body simulation of the globular cluster M4. 
		In addition, \cite{2016MNRAS.462..696Z} compared the radial profiles obtained from  the so-called LIMEPY models to those of simulated snapshots, and studied the evolution of the relevant model parameters in time. \cite{2017MNRAS.470.2736P} compared models and simulated states, by considering a multi-mass generalization of the LIMEPY family of models.
		
		The study presented in this paper makes use of the two-component models introduced by \cite{2016A&A...590A..16D} and is based on the following distribution function:
		
		\begin{equation} \label{fv}
		f^{(\nu)}_{T,i}(E,J) = \begin{cases} A_{i}\exp{\left[-a_i(E-E_t)-d_i \frac{J}{|E-E_t|^{3/4}}\right]} \quad \rm{for} \quad E < E_t 
		\\ 0 \quad \rm{for} \quad E \geq  E_t \end{cases},
		\end{equation}
		
		\noindent where $A_i$, $a_i$, and $d_i$ are positive constants referring to the i-th component. Here, $E = (1/2) v^2 + \Phi(r)$ is the specific energy of a single star subject to a spherically-symmetric mean potential $\Phi(r),$ and $J = |\textbf{r} \times \textbf{v}|$ represents the magnitude of the specific angular momentum. The quantity $E_t = \Phi(r_{t})$ is the truncation potential; the truncation radius $r_{t}$ is assumed to be the same for the two components. 
		
		The index $i$ labels the two species, which we imagine to be light stars of mass $m_1$ and heavy stars of mass $m_2 > m_1$. The total masses of the two components are $M_1$ and $M_2$, respectively. As described in detail by \cite{2016A&A...590A..16D}, the self-consistent models are constructed by integrating the Poisson equation, reduced to a suitable dimensionless form, for the dimensionless potential $\psi = -a_1 (\Phi- E_t)$, to be solved under the boundary conditions $\psi(0)=\Psi = -a_1[\Phi(0) -E_t]$ and vanishing gravitational acceleration at $r=0$. 
		
		From the seven constants $A_1, A_2, a_1, a_2, d_1, d_2, E_t$, we can identify two physical scales, which can be used to match the mass and length scales of a globular cluster, or the corresponding units of a simulated stellar system, and five dimensionless parameters. To reduce the number of free parameters, (1) we set a global mass ratio $M_2/M_1$ to match the mass ratio for the two components, consistent with the way we split the particles of the simulated states into light and heavy particles. Then, (2) we set a central level of partial equipartition by imposing for the central velocity dispersion ratio:
		
		\begin{equation} \label{equi_king_2c}
		\frac{\sigma_1(0)}{\sigma_2(0)} =  \left(\frac{m_1}{m_2}\right)^{-\eta},
		\end{equation}
		
		\noindent where $m_1$ and $m_2$ are the average masses of the particles that are assigned, for a given simulated state, to the light component and to the heavy component, respectively, and $\eta$ is a quantity that follows the notation of \cite{2013MNRAS.435.3272T}. Full energy equipartition would correspond to $\eta = 1/2$, whereas systems characterized by partial energy equipartition present $\eta < 1/2$. In all the cases considered in this paper, a condition of only partial energy equipartition is found in the simulated states, with $\eta \leq 0.27$. Finally, following \cite{2016A&A...590A..16D}, (3) we argue that  $a_1^{1/4} d_1 = a_2^{1/4} d_2$. In conclusion, because of our assumptions (1) - (3), the family is reduced to a two-parameter family of models. The two relevant dimensionless parameters are the concentration of the light component, as represented by $\Psi$, and a second dimensionless parameter $\gamma=a_1 d_1^2/(4 \pi G A_1)$; for a given value of $\Psi$, $\gamma$ has a maximum value, which corresponds to an infinite truncation radius.
		
		In these models, the degree of radial anisotropy increases monotonically from the center to the truncation radius. In particular, the center is isotropic, whereas the outer regions are completely radially anisotropic. Their anisotropy profiles do not present any decrease in the outer parts, as opposed to the case of Michie-King models, for which this feature is taken to represent the effects of the tidal interaction with the host galaxy. 
		
		In order to test the quality of our family of models further, we compared its performance with that of a two-component family of models generated in a similar way (in relation to the choice of $m_1, m_2, M_1, M_2,$ and to the condition of central partial energy equipartition, with the help of the additional parameter $\eta$), but based on the traditional choice of a King distribution function (\citealp{1966AJ.....71...64K}):  
		
		\begin{equation} \label{f_king_2c}
		f^{K}_{i}(E) = \begin{cases} A_{i} \left[ \exp{(-a_{i}E)}-\exp{(-a_i E_{t})} \right] \quad \rm{for} \quad E<E_{t}
		\\ 0 \quad \rm{for} \quad E\geq E_{t},  \end{cases}
		\end{equation}
		
		\noindent where $A_i$ and $a_i$ are positive constants.         As for the $f^{(\nu)}_T$, the quantity $E_t = \Phi(r_{t})$ is the truncation potential; the truncation radius, $r_{t}$, is assumed to be the same for the two components. The family of two-component models is constructed by solving the Poisson equation for the dimensionless gravitational potential $\psi = -a_1 (\Phi- E_t)$, under the boundary conditions $\psi(0)=\Psi = -a_1[\Phi(0) -E_t]$ and vanishing gravitational acceleration at $r=0$. In this case, because the King distribution functions are isotropic, the family of resulting two-component models is characterized by a single dimensionless parameter, the central dimensionless potential $\Psi$ referred to the light component.

		\section{Simulations} \label{sec_simu}
		
		\begin{table*}
			\centering
			\caption{Properties of the simulated states taken from selected snapshots. For each simulated state, the table lists the same properties as in Tab.~ \ref{tab_bianchini_init}, with the addition of the relaxation parameter $n_{rel}$ (following the notation of \citealp{2016MNRAS.458.3644B}). The simulated states are in order of increasing relative relaxation. }
			\label{tab_fin_state}%
			\resizebox{\textwidth}{!}{
				\begin{tabular}{lccccccc}
					\hline \hline
					&$f_{binary} \, (\%)$ & $r_t/r_M$ &N &$M_{tot} \, [M_{\odot}]$ & $r_t \, [pc]$  & $t_{rh} [Gyr]$ & $n_{rel}$ \\ \hline
					Sim 3, 11 Gyr & 5.60  & 6.90  & $4.52 \times 10^5$ & $1.73 \times 10^5$ & 89.11 & 6.61 & 2.5 \\
					Sim 1, 4 Gyr & 3.28  & 14.34 & $4.83 \times 10^5$ & $1.89 \times 10^5$ & 91.83 & 2.41 & 2.8 \\
					Sim 1, 7 Gyr & 3.08  & 12.14 & $4.75 \times 10^5$ & $1.79 \times 10^5$ & 89.68& 3.07 & 4.4 \\
					Sim 1, 11 Gyr & 2.95  & 11.27 & $4.69 \times 10^5$ & $1.73 \times 10^5$ & 88.90 & 3.49 & 8.3 \\
					Sim 6, 4 Gyr & 12.24 & 11.95 & $5.36 \times 10^5$ & $ 2.35 \times 10^5$ &88.29& 1.77 & 10.1 \\
					Sim 5, 4 Gyr & 3.66  & 13.95 & $4.46 \times 10^5$ & $ 1.75 \times 10^5$ &89.29& 1.48 & 14.6 \\
					Sim 6, 7 Gyr & 11.83 & 15.33 & $5.20 \times 10^5$  & $ 2.20 \times 10^5$ & 86.45 & 2.22 & 23.6 \\
					Sim 5, 7 Gyr & 3.54  & 18.21 & $4.32 \times 10^5$ & $ 1.65 \times 10^5$ & 87.67& 1.83 & 64.9 \\
					\hline
				\end{tabular}}%
			\end{table*}%
			We considered the set of Monte Carlo cluster simulations developed and performed by \cite{2010MNRAS.407.1946D} with the Monte Carlo code of \cite{1998MNRAS.298.1239G} (see also \citealp{2013MNRAS.431.2184G} for the description of the method). The simulations include, in addition to a detailed description of the dynamics of the stellar system, the main characteristics of stellar evolution, which produces stellar remnants as a natural outcome.
			By means of a relatively large number of particles (N$= 5\times10^5-2\times10^6$), these simulations provide a realistic description of the long-term evolution of globular clusters, starting with a single stellar population. Selected snapshots taken from these simulations have already been studied as simulated states by \cite{2016MNRAS.458.3644B}  and \cite{2017MNRAS.469.4359B}  to characterize energy equipartition and mass segregation in realistic systems. These authors focused on projected quantities to make useful comparisons with observations. In this paper, we consider intrinsic quantities for a more direct comparison with dynamical models.  
			
			The initial conditions of the simulations under consideration are described in  \cite{2010MNRAS.407.1946D} and \cite{2016MNRAS.458.3644B}. All the simulations include a \cite{2001MNRAS.322..231K} initial mass function with stellar masses ranging from $0.1  \, M_{\odot}$ to $150 \, M_{\odot}$. Different amounts of primordial binaries are considered, either ten percent or fifty percent. The simulations have their initial density  and velocity distributions drawn from a \cite{1911MNRAS..71..460P} isotropic model. An initial cutoff is introduced at 150 pc to mimic the  presence of the tidal field of the host galaxy. This cutoff is not held constant during evolution; it is recalculated at each time step according to the current mass of the cluster, which declines in time because of stellar evolution and dynamical evaporation.	
			Details of the initial conditions of the simulations are summarized in Tab.~\ref{tab_bianchini_init}. The quantities reported here are all intrinsic three-dimensional quantities; in particular, $r_M$ represents the radius of the sphere that includes half of the total mass of the system. The quantity $f_{binary}$ denotes the fraction of the number of particles in binary systems.

			All the systems were evolved for 11 Gyr, and their properties studied at different epochs, $t_{age}=4, 7, 11$ Gyr. In \cite{2016MNRAS.458.3644B}, for each snapshot, the relaxation state of the system is quantified by $n_{rel} = t_{age}/t_{rc}$ , where $t_{rc}$ is the core relaxation time (Eq. 1 in \citealp{2016MNRAS.458.3644B}), following the approach of the \cite{1996yCat.7195....0H} catalog (\citeyear{2010arXiv1012.3224H} edition), defined according to \cite{1993ASPC...50..373D}. Therefore, higher values of $n_{rel}$ correspond to more relaxed stellar systems. The values of $n_{rel}$ for the selected snapshots are taken from Tab.~3 of \cite{2017MNRAS.469.4359B}. The snapshots correspond to pre-core collapse conditions (with respect to the gravothermal catastrophe).
			
			For the present study of energy equipartition and mass segregation, we considered eight simulated states. We focused on the states at 4 Gyr and 7 Gyr of Sim 5 and Sim 6 (for these two simulations, we discarded the snapshots at 11 Gyr, because the associated output files present anomalous features that could not be corrected). We then considered all the available snapshots of Sim 1 as simulated states to study the variation of the relevant properties of the system at different stages of the relaxation process. Finally, we considered the snapshot at 11 Gyr of Sim 3. In this way, we have a sample of simulated states in which $n_{rel}$ varies from 2.5 to 64.9, covering the range between partially relaxed and well dynamically relaxed systems. Table \ref{tab_fin_state} records the same properties as in Tab.~\ref{tab_bianchini_init} for the simulated states considered, with the addition of the relaxation parameter $n_{rel}=t_{age}/t_{rc}$. The snapshots related to Sim 6 present a higher number of particles with respect to the initial conditions as a consequence of the disruption of binaries.
			
			\begin{table}[h]
				\centering
				\caption{Definition of the two components for the simulated states. The various columns list the mean masses of single main sequence stars ($\bar{m}_{MS}$), single giants ($\bar{m}_{giants}$), single remnants ($\bar{m}_{remn}$), and binaries ($\bar{m}_{bin}$). The last column lists the resulting mass of the heavy component. }
				\label{tab_components}%
				\resizebox{\hsize}{!}{
					\begin{tabular}{lcc cc c c}
						\hline \hline
						& \multicolumn{1}{c }{Light stars} &  &  \multicolumn{4}{c}{Heavy stars}   \\  \cline{2-2} \cline{4-7}
						& $\bar{m}_{MS}$ $[M_{\odot}]$ & & $\bar{m}_{giants}$ $[M_{\odot}]$ & $\bar{m}_{remn}$ $[M_{\odot}]$ & $\bar{m}_{bin}$ $[M_{\odot}]$ & $m_{heavy}$ $[M_{\odot}]$ \\ \hline
						Sim 3, 11 Gyr  & 0.30 &  & 0.83  & 0.75  & 0.81  & 0.77   \\
						Sim 1, 4 Gyr  & 0.33 & & 1.12 & 0.86 & 0.83 & 0.86 \\
						Sim 1, 7 Gyr  & 0.31 &  & 0.95  & 0.77  & 0.80  & 0.78  \\
						Sim 1, 11 Gyr  & 0.30 &  & 0.85  & 0.74  & 0.76  & 0.74  \\
						Sim 6, 4 Gyr & 0.34 &  & 1.12  & 0.82  & 0.78  & 0.80   \\
						Sim 5, 4 Gyr & 0.33 &  & 1.12  & 0.81  & 0.77  & 0.81  \\
						Sim 6, 7 Gyr  & 0.32 &  & 0.94  & 0.77  & 0.73  & 0.75 \\
						Sim 5, 7 Gyr  & 0.32 &  & 0.94  & 0.76  & 0.73  & 0.75   \\ 
						\hline
					\end{tabular}}%
				\end{table}%
				
				\subsection{Mass segregation in the simulated states}
\begin{figure}[h]
					\includegraphics[width=\hsize]{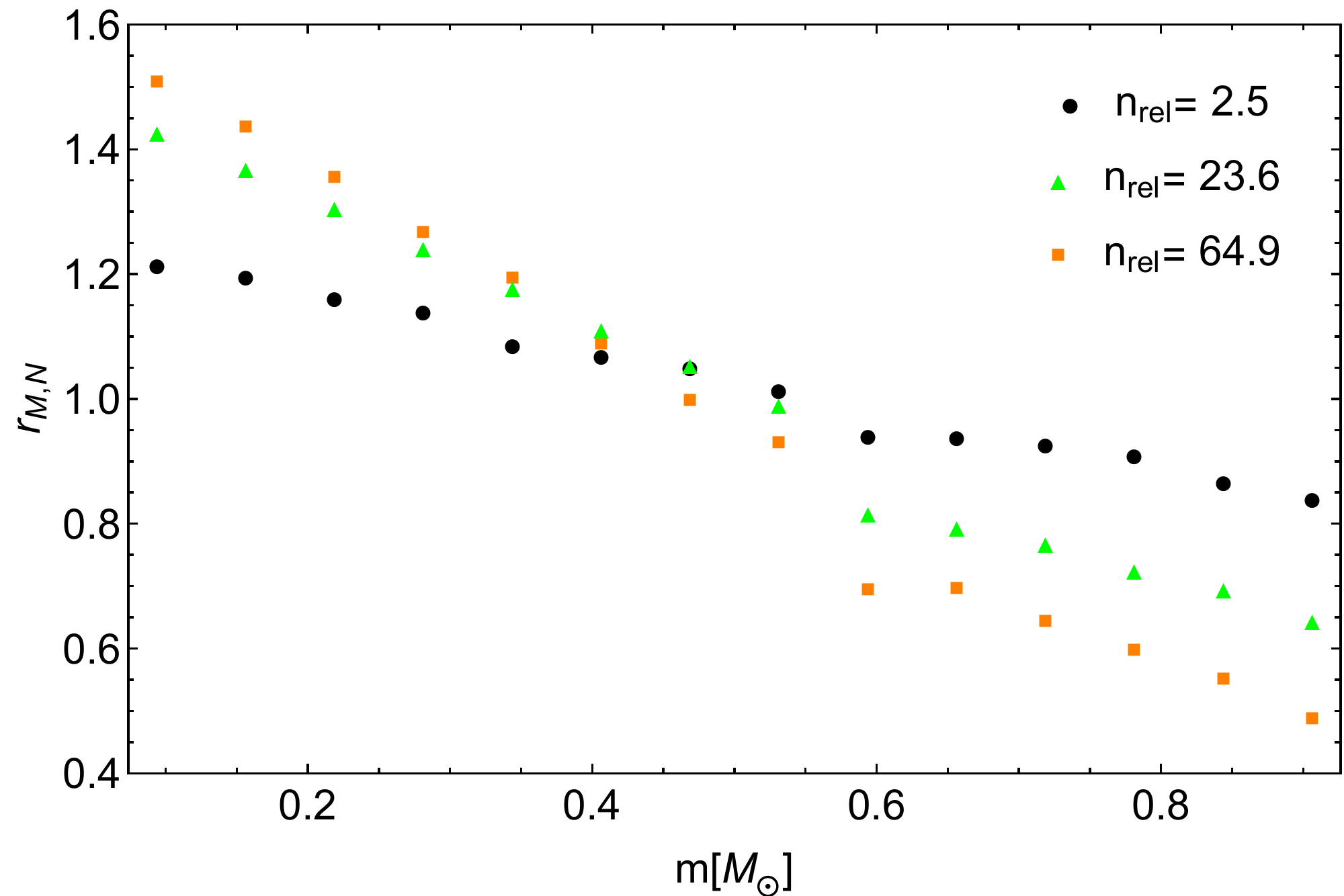}  
					\caption{Variation of $r_{M,N}$, that is the half-mass radius normalized to global half-mass-radius of each simulation, with stellar mass. The simulations considered, Sim 3 at 11 Gyr (black circles), Sim 6 at 7 Gyr (green triangles), and Sim 5 at 7 Gyr (orange squares), show a decreasing trend with mass. The fact that also the least relaxed simulated state (with $n_{rel}=2.5$) shows mass segregation suggests that this process sets in efficiently after very few relaxation times. }\label{fig_half_mass_radii}
				\end{figure}

				The presence of mass segregation can be observed by considering the variation of the half-mass radius for stars with different masses as a function of mass (see also Fig. 9 in \citealp{2016A&A...590A..16D} and related discussion). For a mass-segregated system, we expect a decreasing trend with stellar mass. In Fig. \ref{fig_half_mass_radii}, this trend is also present for the least relaxed system of our sample, Sim 3 at 11 Gyr. More relaxed simulated states (Sim 6 at 7 Gyr and Sim 5 at 7 Gyr) exhibit a steeper slope, thus showing a higher degree of mass segregation.
				We also note that the gradient is small in all the half-mass radius profiles  at about $0.6 \, M_{\odot}$: this is the typical mass at which white dwarfs form (given the initial mass function, the initial-final mass relation and the age of the selected snapshots), in many cases after undergoing a severe mass loss. If, after loosing a great quantity of mass, white dwarfs do not have time to relax dynamically, they will be characterized by the same phase-space properties as objects with their original mass and, as a consequence, they will exhibit a lower half-mass radius. A similar feature was noted for the level of energy equipartition in \cite{2016MNRAS.458.3644B} (see their Fig. 2).

				\subsection{Definition of the two components} \label{sec_2c_definition}
				
				To compare the relevant properties of the systems under consideration to the models introduced in Sect.~\ref{sec_2cmodels}, we need to define the two components for each simulated state. As anticipated, the definition of the components is based on the mass of the stars. 
				
				First, we divided the stars of the simulated states into four classes: single main sequence stars, single giant stars, single remnants (white dwarfs, neutron stars, and black holes), and binaries. In Tab.~\ref{tab_components}, we give the mean mass of the four classes of stars. The mean masses of giants, remnants, and binaries are always more than twice the mean mass of main sequence stars. We thus identified the main sequence stars with the light component (with mass $m_{light}=\bar{m}_{MS}$) and combined the other classes into the heavy component (with $m_{heavy}$ equal to the the mean mass of stars belonging to these three classes). Our choice for identifying the two components was preferred with respect to other options, such as a simple mass cut, because it is physically motivated and may turn out to be useful when the models will be considered for diagnostics of observed clusters.
				
\begin{figure}[h]
					\centering
					\includegraphics[width=\hsize]{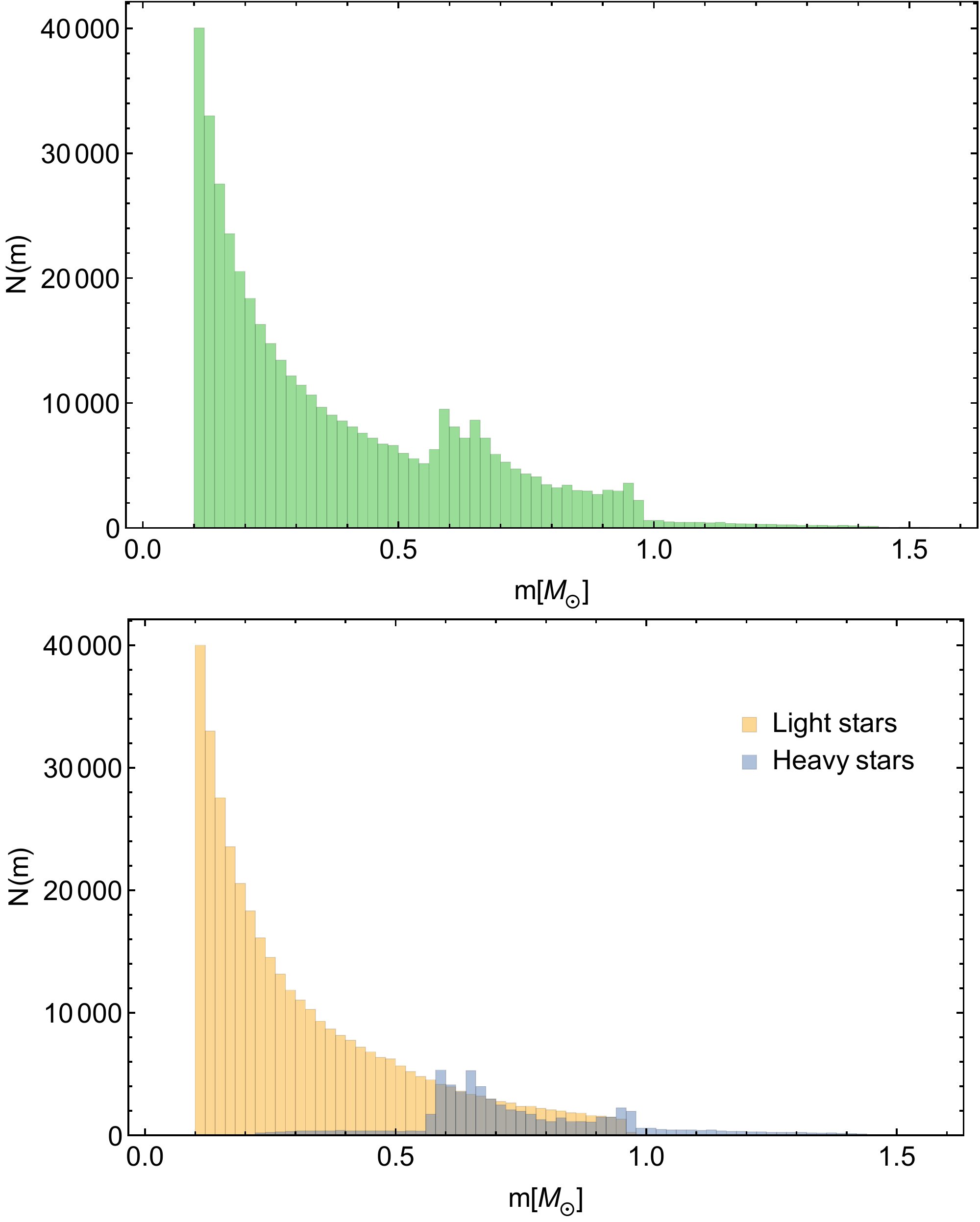}  
					\caption{Mass spectrum of Sim 5 at 7 Gyr constructed by considering all the stars of the simulated state (upper panel) and by separating the light and the heavy stars (lower panel). The presence of a secondary peak in the total mass spectrum is due to the formation of white dwarfs, with a typical mass of $0.6 \, M_{\odot}$.}\label{fig_mass_spectrum}
				\end{figure}

				In Fig. \ref{fig_mass_spectrum}, we illustrate the mass spectrum for Sim 5 at 7 Gyr.
				The spectrum obtained considering all the stars shows a general decreasing trend, with a secondary peak at about $0.6 \, M_{\odot}$, which is the typical mass at which white dwarfs form. For this reason,  by separating the spectra of the two components, we observe that the spectrum of the heavy component has a maximum at this value of mass, whereas the mass spectrum of the light component is monotonically decreasing.

				\subsection{Intrinsic profiles}
				
\begin{figure}[h]
					\includegraphics[width=\hsize]{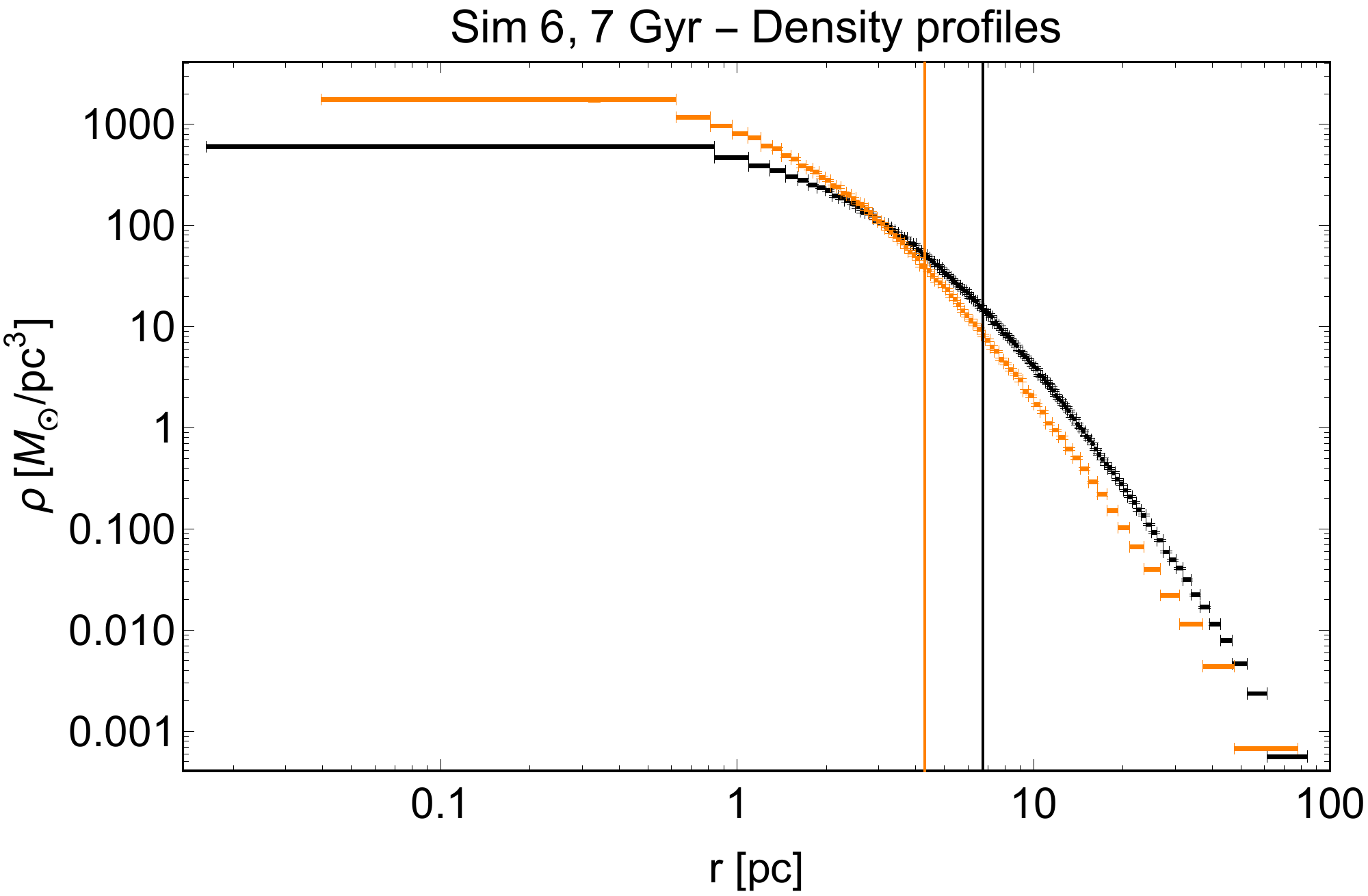}  
					\includegraphics[width=\hsize]{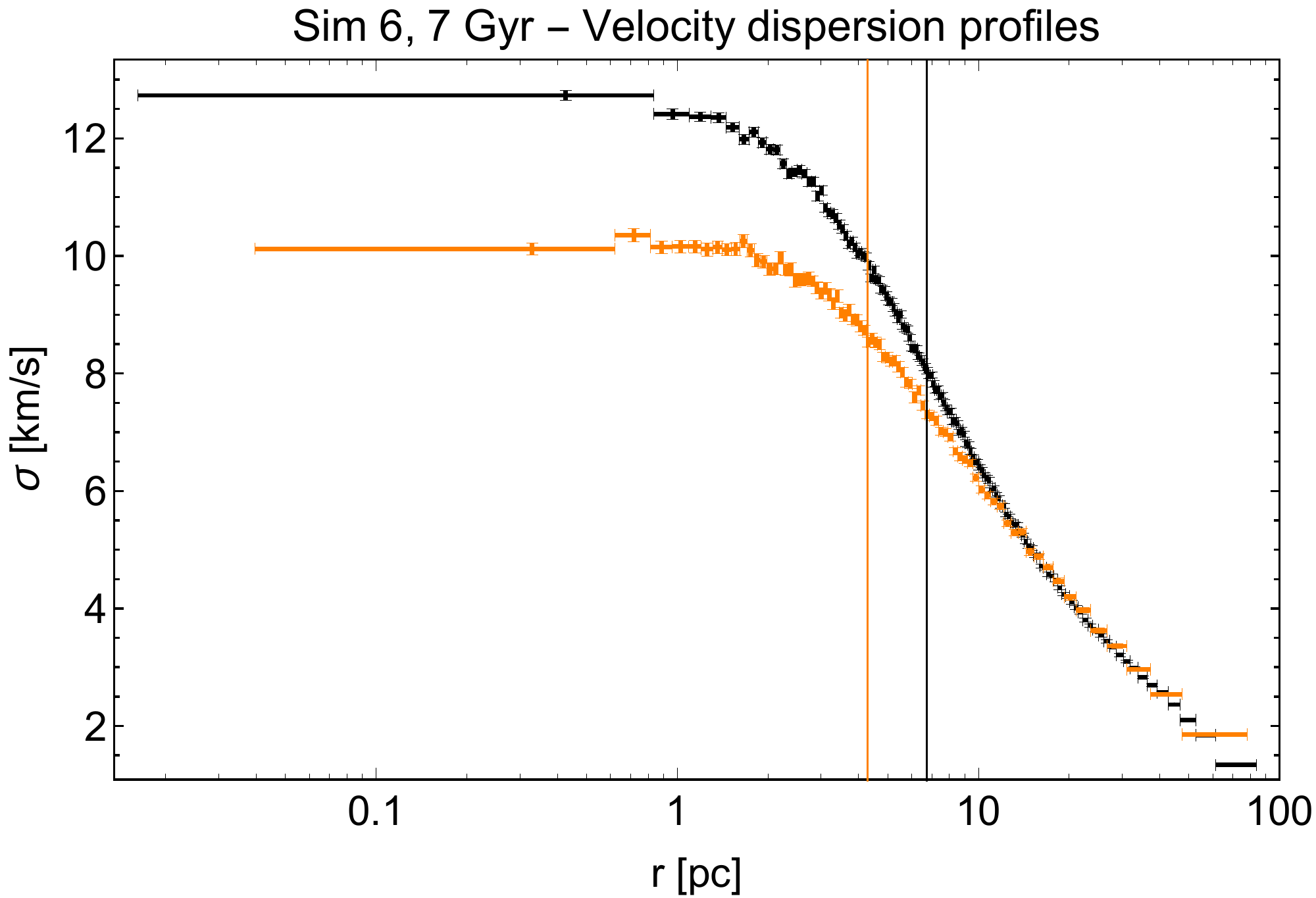} 
					\caption{Density (upper panel) and velocity dispersion (lower panel) profiles for light (black) and heavy (orange) component of Sim 6 at 7 Gyr. Uncertainties on the profile values are obtained by means of a bootstrap resampling. The differences between the two density profiles indicate the presence of mass segregation, whereas the differences between the two velocity dispersion profiles may be interpreted as due to the effects of partial energy equipartition. The vertical lines indicate the half-mass radius of the component under consideration.}
					\label{profiles_sim6_7gyr}%
				\end{figure}

				From the output of the simulations, we constructed density and velocity dispersion profiles for the entire system and, separately, for each of the two components. The profiles were constructed by dividing the system into radial shells with a constant number of stars. The radial errors were evaluated as the width of each shell. In turn, the density and velocity dispersion uncertainties were determined by means of a bootstrap resampling (\citealp{efron1986}). The uncertainties obtained in this way are very small, typically of the order of 1\%. Figure \ref{profiles_sim6_7gyr} illustrates the density and the velocity dispersion profiles for the two components of Sim 6 at 7 Gyr. The quantity $\sigma$ denotes the total, local velocity dispersion $\sigma = \sqrt{\sigma_{rr}^2 + \sigma_{\theta \theta}^2 + \sigma_{\phi \phi}^2}$.
				
				A local measure of the pressure anisotropy is given by the function $\alpha(r)$, defined as:
				
				\begin{equation} \label{alpha_def}
				\alpha(r) = 2 - \frac{\sigma^2_t(r)}{\sigma^2_r(r)},
				\end{equation}
				
				\noindent where $\sigma^2_t = \sigma_{\theta \theta}^2 + \sigma_{\phi \phi}^2$ and $\sigma^2_r = \sigma_{rr}^2$ are the tangential and radial velocity dispersions (squared), respectively.  Isotropy corresponds to $\alpha=0$, whereas $\alpha=2$ represents a condition of complete radial anisotropy.
				
\begin{figure}[h]
					\includegraphics[width=\hsize]{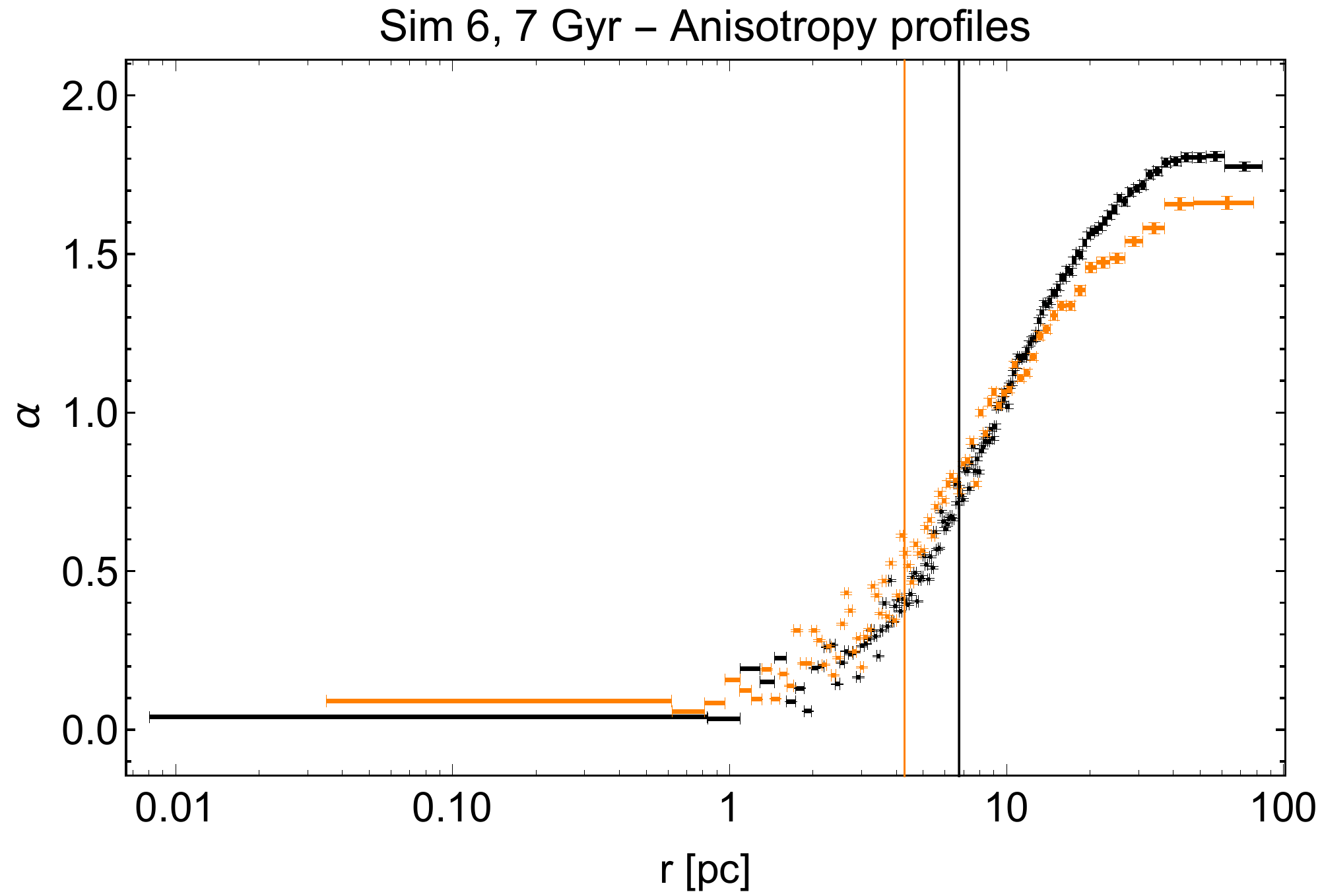}  
					\caption{Anisotropy profiles for light (black) and heavy (orange) component of Sim 6 at 7 Gyr. Uncertainties on the profile values are obtained by propagating the errors on the different velocity dispersions. The vertical lines indicate the half-mass radius of the component under consideration.}
					\label{anisotropy_profiles_sim6_7gyr}%
				\end{figure}

				In Fig.~\ref{anisotropy_profiles_sim6_7gyr}, we show the anisotropy profiles for the two components of Sim 6 at 7 Gyr.
				\noindent The simulated state is characterized by a monotonically increasing anisotropy profile. Even though the system under consideration has been initialized with an isotropic distribution of velocities, the slow cumulative effects of relaxation processes have led the system toward a velocity distribution that resembles that generated by collisionless violent relaxation (which is the physical basis under which the $f^{(\nu)}$ models were originally constructed), with an isotropic core and a radially-biased anisotropic envelope.        
				
				\subsection{Basic parameters for a comparison with two-component models}
				
				\begin{table}[ht]
					\centering
					\caption{Basic parameters of the simulated states for a comparison with two-component models: the heavy to light mean mass ratio, the light to heavy total mass ratio, and the equipartition parameter.  The last two columns list the value of $\eta_{eq}$, that is the values of $\eta$ obtained by means of Eq. (4) of \cite{2016MNRAS.458.3644B}, and of the Spitzer parameter $S = (M_{heavy}/M_{light}) (m_{heavy}/m_{light})^{3/2}$. }
					\label{basic_parameters}%
					\resizebox{\hsize}{!}{
						\begin{tabular}{lcc cc c c c c c c c c}
							\hline \hline
							
							& $m_{heavy}/m_{light}$ & $M_{light}/M_{heavy}$ & $\eta$ & $\eta_{eq}$ & $S$ ($S_{max}=0.16$) \\ \hline 
							Sim 3, 11 Gyr  & 2.57 &  1.78 & 0.191 & 0.180 & 2.31  \\
							Sim 1, 4 Gyr  & 2.61 & 2.94 & 0.204 & 0.161 & 1.43 \\ 
							Sim 1, 7 Gyr  & 2.50 & 2.54 &  0.212 & 0.172 & 1.56   \\
							Sim 1, 11 Gyr  & 2.46 & 2.29  & 0.232 & 0.208 & 1.68   \\
							Sim 6, 4 Gyr & 2.37 & 1.57 & 0.230 & 0.231 & 2.33   \\
							Sim 5, 4 Gyr & 2.42 & 2.94 & 0.259 & 0.260 & 1.28  \\
							Sim 6, 7 Gyr  & 2.34 & 1.40 &  0.269 & 0.233 & 2.55   \\ 
							Sim 5, 7 Gyr  & 2.38 & 2.43 & 0.270 & 0.269 & 1.52  \\ \hline
						\end{tabular}}%
					\end{table}%
					
					In Tab.~\ref{basic_parameters}, we list the parameters used to reduce the number of free constants in the two-component models defined in Sect. ~\ref{sec_2cmodels}: the ratio $(m_2/m_1)$ of the single masses of the two components, the ratio $M_2/M_1$ of the total masses of the two components, and the parameter $\eta$, which quantifies the degree of central energy equipartition. The ratio $m_{heavy}/m_{light}$ is very similar in all the simulated states. Instead, the ratio $M_{light}/M_{heavy}$ decreases with the age of the cluster as more and more remnants are produced (and, possibly, because of evaporation of low mass stars). In addition, $M_{light}/M_{heavy}$ is lower for the two snapshots of Sim 6, because of the high number of binaries that the system was initialized with. In passing, we note that the components as defined above would violate the Spitzer criterion (\citealp{1969ApJ...158L.139S}). In fact, the value of $S = (M_{heavy}/M_{light})(m_{heavy}/m_{light})^{3/2}$, which (according to \citealp{1969ApJ...158L.139S}) should be less than $S_{max}=0.16$ for a system in thermal and virial equilibrium, is greater by about an order of magnitude in the simulated states under consideration. The violation of the Spitzer criterion is consistent with the fact that a condition of total energy equipartition is not fulfilled by the systems under consideration.
					
					The equipartition parameter $\eta$ was determined by evaluating  the ratio of the velocity dispersions of the two components at the center of the simulated states and by equating $\sigma_{light}(0)/\sigma_{heavy}(0) = \left(m_{light}/m_{heavy}\right)^{-\eta}$, where $\sigma(0)$ indicates the velocity dispersion of the central bin. As indicated in Tab.~\ref{tab_components}, the most relaxed systems present the highest values of $\eta$, confirming that relaxation processes are driving them toward a condition of central energy equipartition; however, values close to $\eta=0.5$ are never attained. Column 10 of Tab.~\ref{tab_components} lists the values of $\eta_{eq}$, which is the value of $\eta$ calculated directly from the equipartition parameter $m_{eq}$ by means of Eq. (4) of \cite{2016MNRAS.458.3644B}; the values of mass considered are the mean masses of the stars in the first radial bin. The values obtained for $\eta_{eq}$ are fairly close to those used in this paper.
					
\begin{figure}[h]
						\centering
						\includegraphics[width=\hsize]{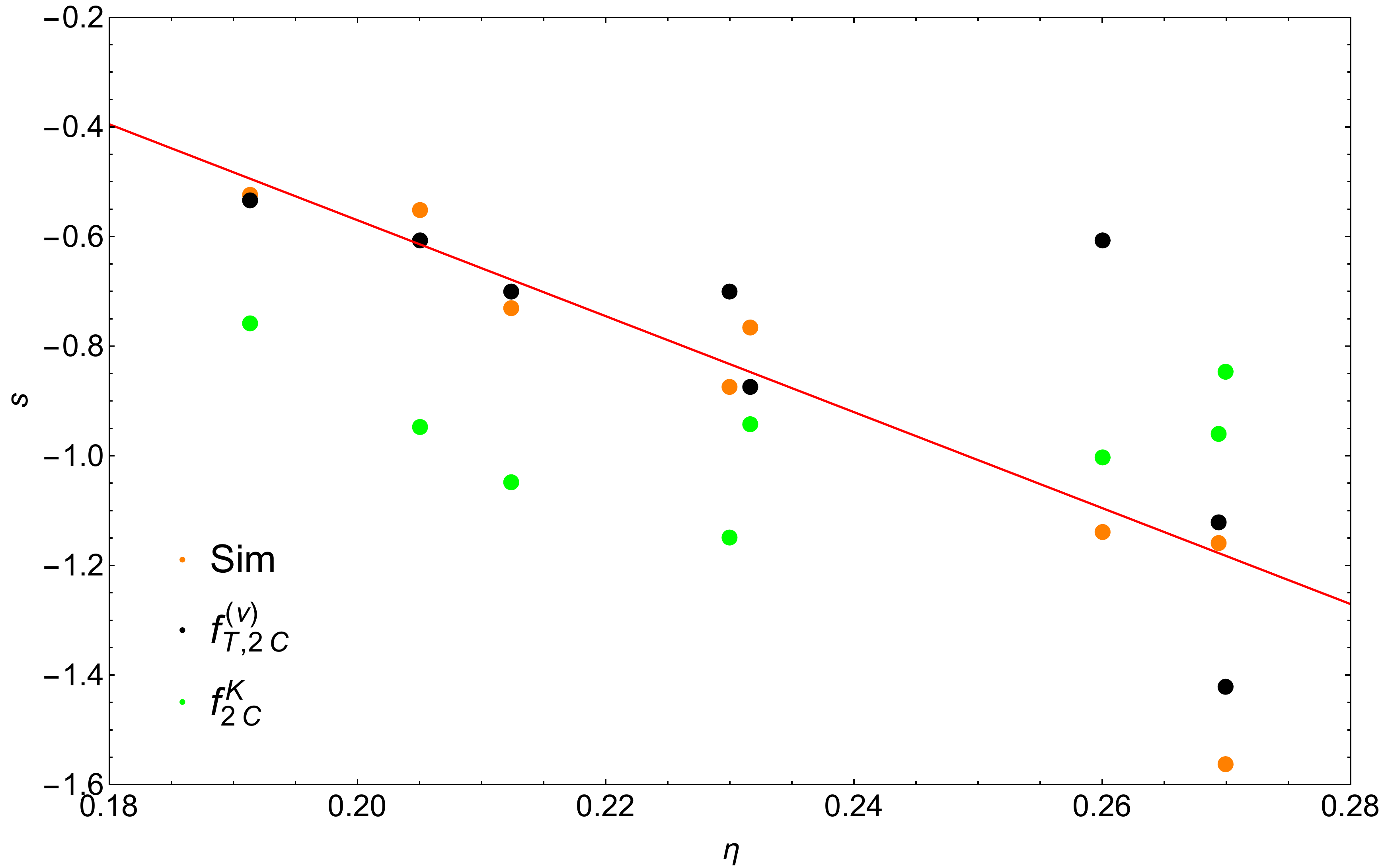}  
						\caption{Relation between slope of the half-mass radius profiles
							and equipartition parameter $\eta$ for simulated states (orange): a linear relation
							(red line) is found between the two parameters. The linear relation shows that clusters characterized by a higher level of energy equipartition display a higher  level of mass segregation. To be comprehensive, we also show the corresponding points for which the half-mass radii are derived from the best-fit two-component $f_T^{(\nu)}$ models as black dots, and those obtained from the best-fit two-component King models (to be described in the next section) as green dots. There is a better correspondence between simulated states and the $f_T^{(\nu)}$ models with respect to the King models.} 
						\label{slope_eta_relation}%
					\end{figure}

					In Fig. \ref{slope_eta_relation}, we show the relation between the degree of mass segregation and the equipartition parameter $\eta$. The degree of mass segregation is quantified by \textit{s}, that is the slope of the half-mass radius profile as a function of stellar mass (see Fig. \ref{fig_half_mass_radii}). This is calculated by including all the components defined in Sect. \ref{sec_2c_definition}  (white dwarfs that underwent a severe mass loss are also included). A linear relation is found between the values of \textit{s} and $\eta$ in different simulated states (orange dots), with the exception of the most relaxed simulated state, Sim 5 at 7 Gyr. This linear relation is well reproduced by best-fit two-component $f_T^{(\nu)}$ models (black dots, see Sect. \ref{sec_fitting_2c}). 
					
					\section{A description of the simulated states by two-component models}
					
					\begin{table*}[h]
						\centering
						\caption{Best-fit parameters for two-component models.}
						\resizebox{\textwidth}{!}{
							\begin{tabular}{l c c c c c c c c c c c c c c}
								\hline \hline
								&       & \multicolumn{4}{c}{King models} & & \multicolumn{8}{c}{$f^{(\nu)}_T$ models}   \\  \cline{2-7} \cline{9-15}
								& $\Psi$ & $\tilde{\chi}^2_{\rho_1}$ &  $\tilde{\chi}^2_{\sigma_1}$ &  $\tilde{\chi}^2_{\rho_2}$ & $\tilde{\chi}^2_{\sigma_2}$ &  $\tilde{\chi}^2_{tot}$ & &$\Psi$ &  $\gamma$ &  $\tilde{\chi}^2_{\rho_1}$  &  $\tilde{\chi}^2_{\sigma_1}$ & $\tilde{\chi}^2_{\rho_2}$ & $\tilde{\chi}^2_{\sigma_2}$ &  $\tilde{\chi}^2_{tot}$  \\ \hline
								Sim 3, 11 Gyr & $4.700 \pm 0.002$ & 68.44  & 99.95 & 31.12 & 36.24 & 67.09 &  & $3.85 \pm 0.01 $& $28.4 \pm 0.1$ & 18.90  & 21.90 & 5.25 & 15.58 & 16.82 \\
								Sim 1, 4 Gyr & $4.717 \pm 0.005$ & 102.35 & 31.62 & 577.52 & 196.80 & 286.93 & & $4.03 \pm 0.01 $ & $56.5 \pm 0.2$ & 7.05  & 4.59  & 44.38  & 16.49  & 21.16   \\
								Sim 1, 7 Gyr &  $4.751 \pm 0.004 $ & 103.03 & 26.89 & 465.23 & 171.13 & 233.75 & & $4.14 \pm 0.01$ & $54.0 \pm 0.2$ & 6.94 & 18.54  & 29.68  & 22.76 & 21.32     \\
								Sim 1, 11 Gyr & $5.165 \pm 0.003$ & 104.06 & 357.93  & 53.56 & 162.53 & 195.59 & & $4.467 \pm 0.002$ & $47.8 \pm 0.2$& 13.58  & 18.05 & 7.73 & 15.89 & 14.46  \\
								Sim 6, 4 Gyr & $4.651 \pm 0.003$ & 156.59 & 689.87  & 45.19 & 219.91 & 312.14 & & $4.103 \pm 0.009$ & $67.4 \pm 0.2$ & 9.65 & 20.82 & 8.15 & 8.81 & 12.52\\
								Sim 5, 4 Gyr & $5.651 \pm       0.004$ &        160.24  & 712.63 &  86.34 &    157.21 &        358.92 &  & $4.708 \pm 0.001$ & $67.0 \pm 0.2$ & 7.02  &   30.92 & 8.11 & 13.22 & 16.70   \\
								Sim 6, 7 Gyr & $4.855 \pm 0.004$ & 164.17  & 671.29 & 67.80 & 217.85 & 312.82 &  & $4.115 \pm 0.003$ & $56.6 \pm 0.2$ & 3.68 & 21.35  & 7.05 & 6.59 & 10.23  \\
								Sim 5, 7 Gyr & $5.680 \pm 0.003$ & 201.66 & 735.86  & 49.67 & 150.21 & 364.36 & & $5.46  \pm 0.02$ & $49.0 \pm 0.2$ & 7.13& 20.76  & 16.34 & 13.05 & 13.89  \\
								\hline
							\end{tabular}}%
							\label{bestfit_2c}%
						\end{table*}%
						
						We performed a combined chi-squared analysis on the density and velocity dispersion profiles, following a procedure very similar to that outlined in \cite{2012A&A...539A..65Z}. In the present analysis, we decided to minimize a combined chi-squared function, defined as the sum of the  two-component density and velocity dispersion chi-squared. The two components to be modeled as light and heavy stars are those defined in Sect.~\ref{sec_2c_definition}. In order to deal with a small number of free parameters, differently from the fits reported in \cite{2012A&A...539A..65Z}, the physical scales of the models were set by equating the length and the mass scales of the models to those of the simulated states, that are known a priori. 
						
						We wish to emphasize that this is intended as a formal analysis with the objective of determining whether an idealized two-component model is able to give a reasonable description of the simulated states under consideration. As such, this is only a prerequisite for a follow-up analysis in which we intend to apply and test the models in order to judge their accuracy as diagnostic tools.

						\subsection{Fit by two-component models} \label{sec_fitting_2c}
						
\begin{figure}[]
							\centering
							\includegraphics[width=\hsize]{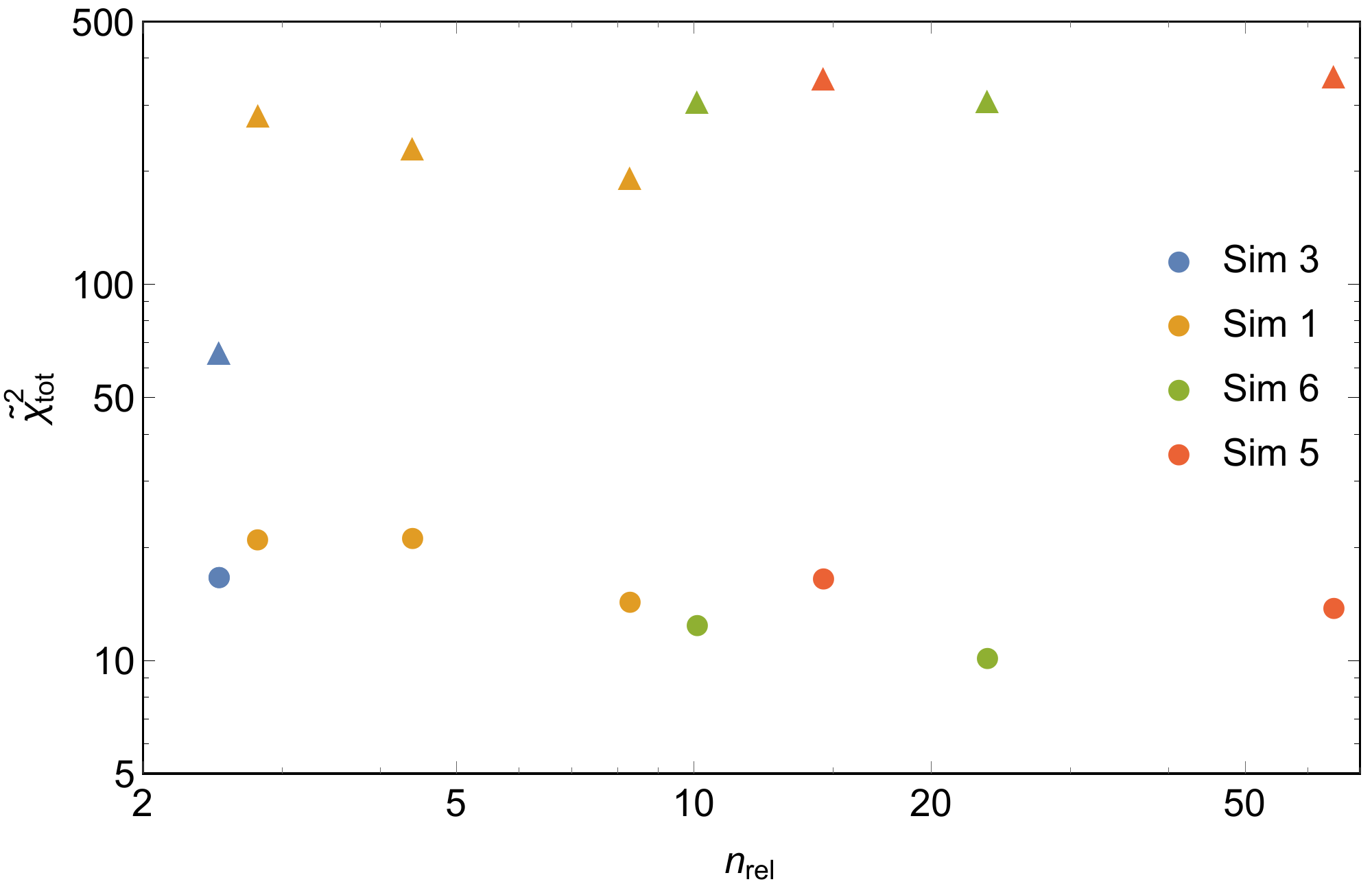}  
							\caption{Values of $\tilde{\chi}^2_{tot}$ (plotted against $n_{rel}$) associated with two-component King models (triangles) and $f_T^{(\nu)}$ models (circles). The plot shows that the latter models perform far better than the former models.}\label{global_comparison}
						\end{figure}

						The best-fit values of the dimensionless parameters (with their formal errors) and the reduced chi-squared for two-component models are listed in Tab.~\ref{bestfit_2c}.
						The total reduced chi-squared was calculated by dividing the sum of the chi-squared of the individual components by the total number of degrees of freedom, which is the number of data points minus the number of free parameters. 
						Once the best-fit model was found, we estimated the single reduced chi-squared separately (by dividing the density and velocity chi-squared by the respective number of degrees of freedom) to evaluate the quality of the single comparisons. We recall that the $f^{(\nu)}_T$ models are characterized by two dimensionless parameters, $\Psi$ and $\gamma$, whereas the King models are characterized by only one parameter, $\Psi$; in both cases, the parameters are referred to the light component. 
						The set of simulated states is listed in order of increasing relaxation. Figure \ref{global_comparison} shows the total discrepancies (quantified by $\tilde{\chi}^2_{tot}$) plotted against $n_{rel}$ for the two models considered. The King models perform systematically worse with respect to the $f_T^{(\nu)}$ models. There seems to be no systematic trend between the relevant dimensionless parameters and the relaxation state.
						From inspection of Tab. \ref{bestfit_2c} and of Fig. \ref{global_comparison}, we can draw some conclusions, which are then confirmed by considering the plots of the relevant profiles.
						
						\begin{figure*}
							\centering
							\includegraphics[width=1\textwidth]{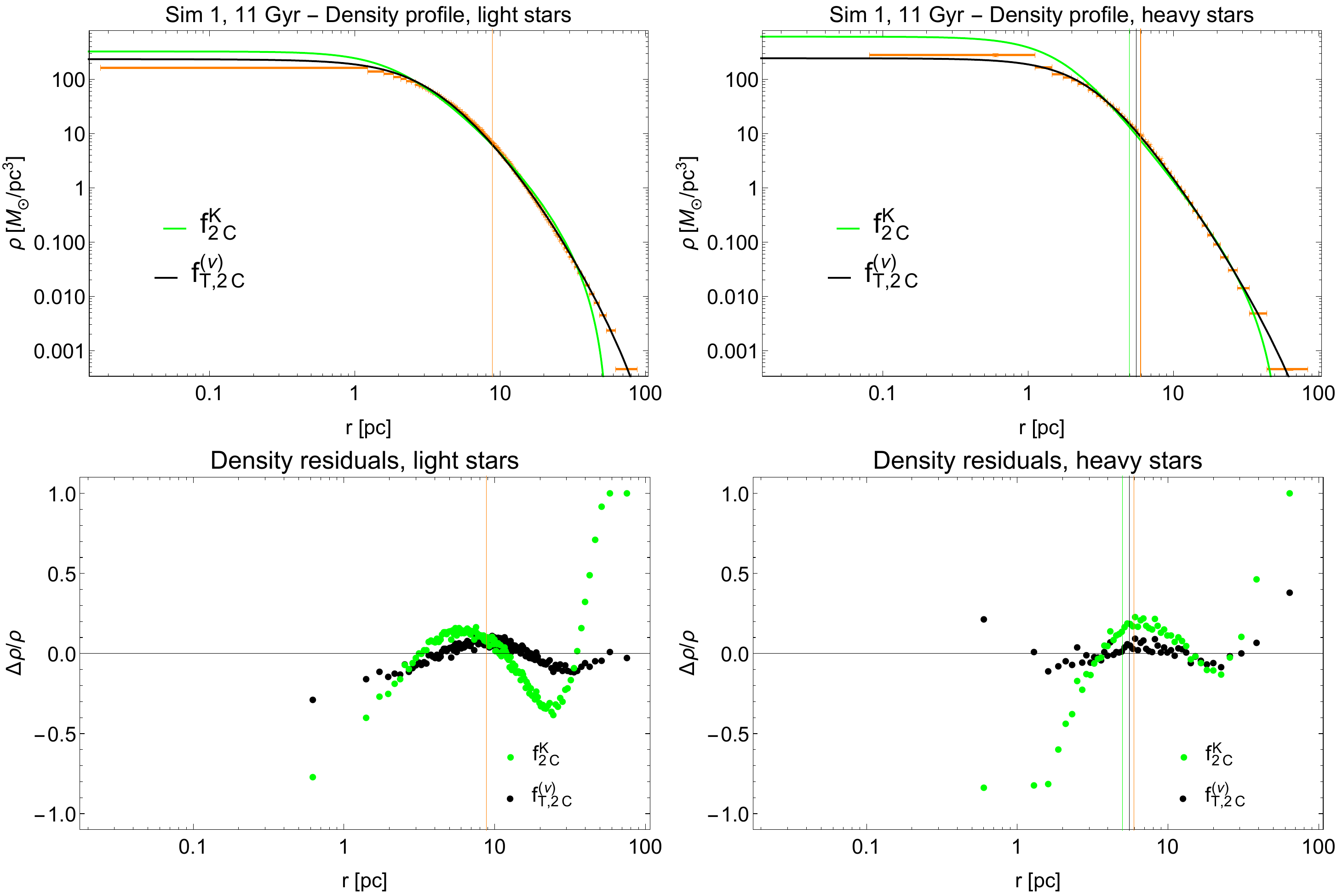} 
							\caption{Best-fit profiles and residuals for Sim 1 at 11 Gyr for two-component King models (green) and for two-component $f^{(\nu)}_T$ models (black). Upper panels: density profile for light component (left) and for heavy component (right). Lower panels: density residuals for light component (left) and for heavy component  (right).  Vertical lines represent the half-mass radius of the component under consideration in the simulated state and in the models. }\label{fig_2c_sim1_11_density}
						\end{figure*}

						For the King models, the values of  $\tilde{\chi}_{tot}^2$ are very high. Indeed, there is little agreement  (in the outer regions discrepancies are always of \(\approx \! \!100 \%\)) between the model and the simulated system. In some cases, the density of the heavy component exhibits smaller values of the reduced chi-squared; the best result appears to be obtained for the least relaxed state, Sim 3 at 11 Gyr (but we should recall that the systems were all initialized with isotropic initial conditions). In general, the density profiles of the two-component King models present a truncation that is too sharp. This feature was already noted in observed data (e.g., \citealp{2005ApJS..161..304M}) and N-body simulations (e.g., \citealp{2016MNRAS.462..696Z}). In addition, the central densities predicted by these models are too high with respect to those of the simulated states. The high values of $\tilde{\chi}_{tot}^2$ are largely related to the sharp truncation of the models, which causes great discrepancies (up to 100\%) in the outer regions. The kinematic comparison gives the worst results, especially for the light component. The velocity dispersion profiles, as noted also for one-component models (see Appendix \ref{App.A}), are too flat in the central regions, and decrease too rapidly in the outermost parts of the system. The large kinematic discrepancies are probably due to the fact that King models are isotropic, whereas the simulated states are radially-biased anisotropic systems.
						
						The two-component $f^{(\nu)}_T$ models appear to offer a better representation of the simulated states, judging from the associated values of the $\tilde{\chi}_{tot}^2$. In fact, they are found to give a good description of the density profiles for the two components of the simulated states  (generally the maximum residual does not exceed 20\%), both in the outermost regions, thanks to a milder truncation, and in the central parts. In addition, they are able to reproduce (with discrepancies of less than 10\%) the central peak in the velocity dispersion profiles, especially for more relaxed systems.

\begin{figure*}
							\centering
							\includegraphics[width=1\textwidth]{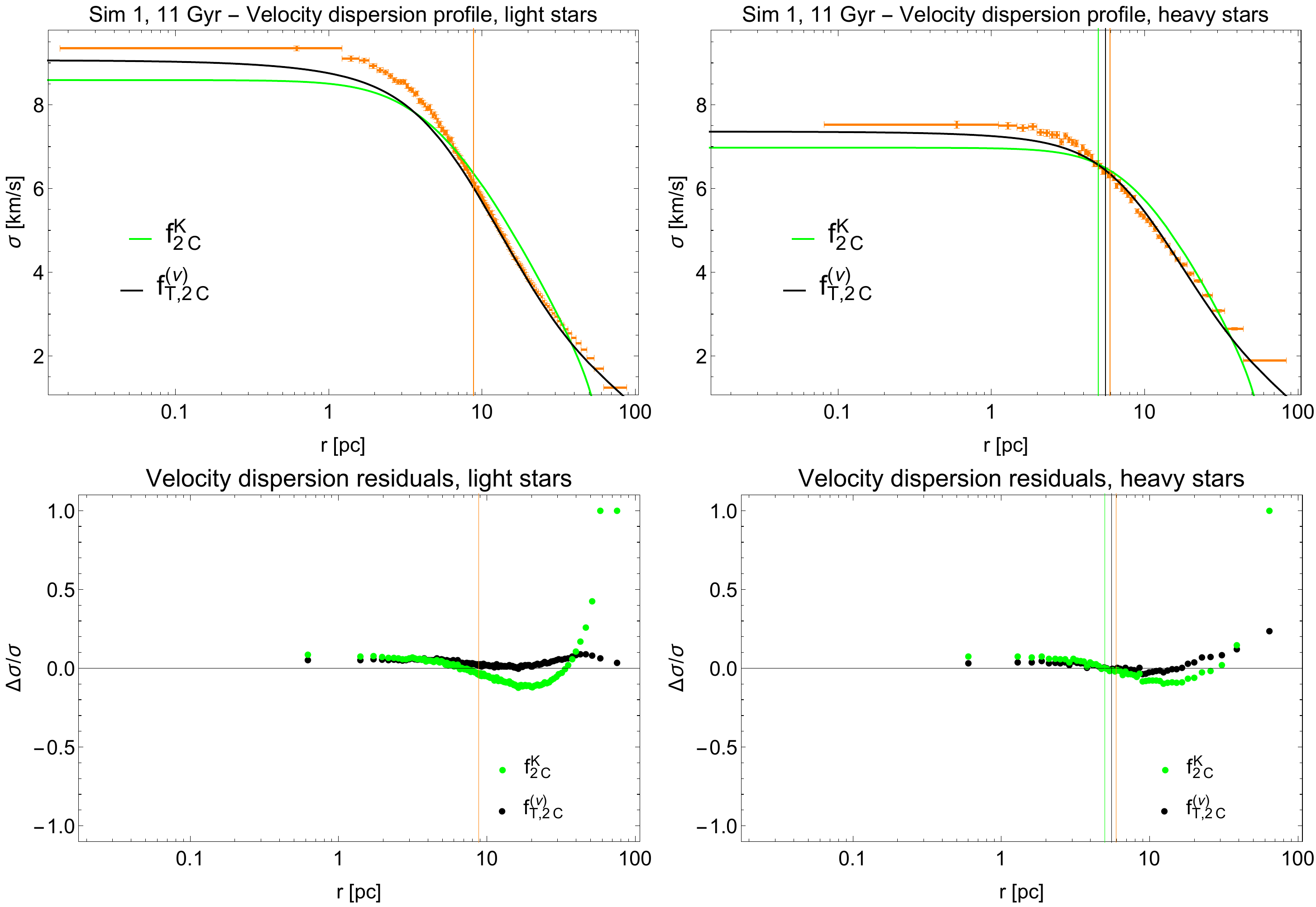}  
							\caption{ Best-fit profiles and residuals for Sim 1 at 11 Gyr for two-component King models (green) and for two-component $f^{(\nu)}_T$ models (black). Upper panels: velocity dispersion profile for light component (left) and for heavy component (right). Lower panels: velocity dispersion residuals for light component (left) and for heavy component  (right). The vertical lines represent the half-mass radius of the component under consideration in the simulated state and in the models.  }\label{fig_2c_sim1_11_velocity}
							
						\end{figure*}

\begin{figure*}
							\centering
							\includegraphics[width=1\textwidth]{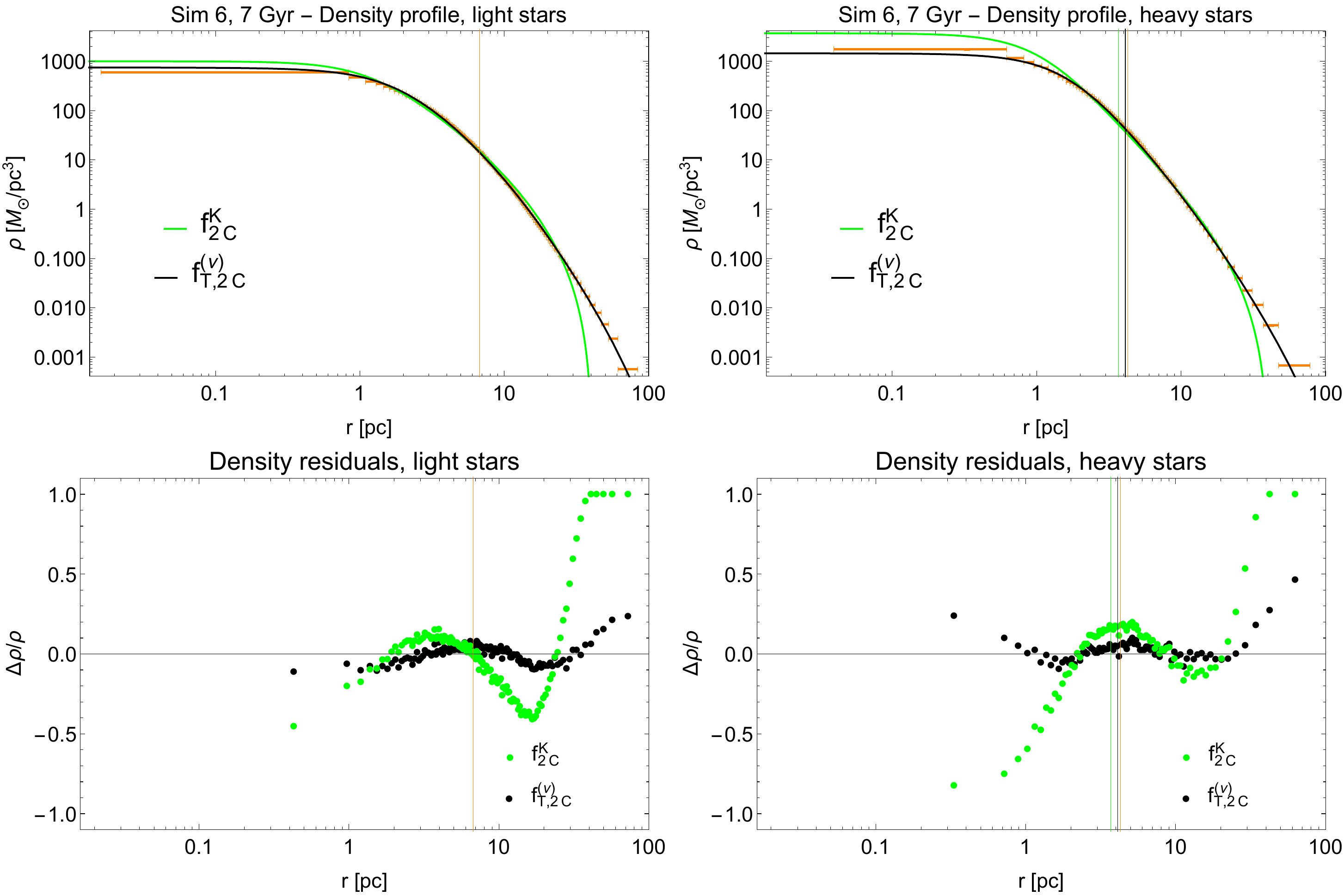} 
							\caption{ Best-fit profiles and residuals for Sim 6 at 7 Gyr for two-component King models (green) and for two-component $f^{(\nu)}_T$ models (black). Upper panels: density profile for light component (left) and for heavy component (right). Lower panels: density residuals for light component (left) and for heavy component  (right). The vertical lines represent the half-mass radius of the component under consideration in the simulated state and in the models.}\label{fig_2c_sim6_7_density}
						\end{figure*}

\begin{figure*}[!htb]
							\centering
							\includegraphics[width=\hsize]{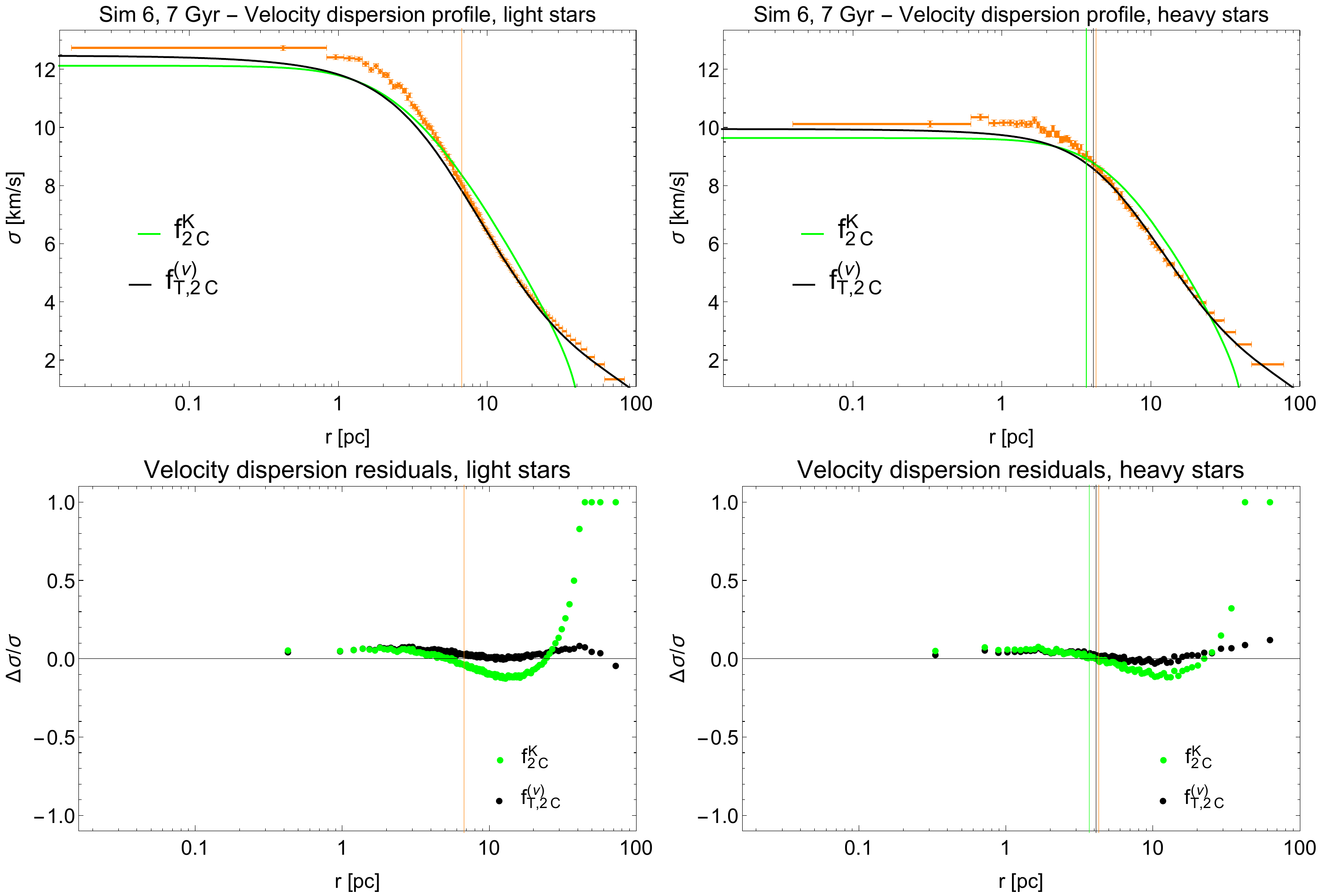}  
							\caption{ Best-fit profiles and residuals for Sim 6 at 7 Gyr for two-component King models (green) and for two-component $f^{(\nu)}_T$ models (black). Upper panels: velocity dispersion profile for light component (left) and for heavy component (right). Lower panels: velocity dispersion residuals for light component (left) and for heavy component  (right). The vertical lines represent the half-mass radius of the component under consideration in the simulated state and in the models. }\label{fig_2c_sim6_7_velocity}
						\end{figure*}

						As an illustration of the above conclusions for simulated states under conditions of intermediate and advanced relaxation, we show, in detail, the best-fit profiles for two cases. In Fig.~\ref{fig_2c_sim1_11_density} and Fig.~\ref{fig_2c_sim1_11_velocity}, we show the best-fit density and velocity dispersion profiles for the two components of Sim 1 at 11 Gyr, and in Fig.~\ref{fig_2c_sim6_7_density} and Fig.~ \ref{fig_2c_sim6_7_velocity}, we show the best-fit profiles for Sim 6 at 7 Gyr. The vertical lines represent the half-mass radii (referred to each component) for the simulated state (red line), for the best-fit $f_T^{(\nu)}$ model (black solid line), and for the best-fit King model (black dashed line). For the light component, the half-mass radii coincide because the scale length of the models is set by equating its light component half-mass radius to that of the simulated state.

						
						\begin{figure*}[!hb]
							\centering
							\includegraphics[width=1\textwidth]{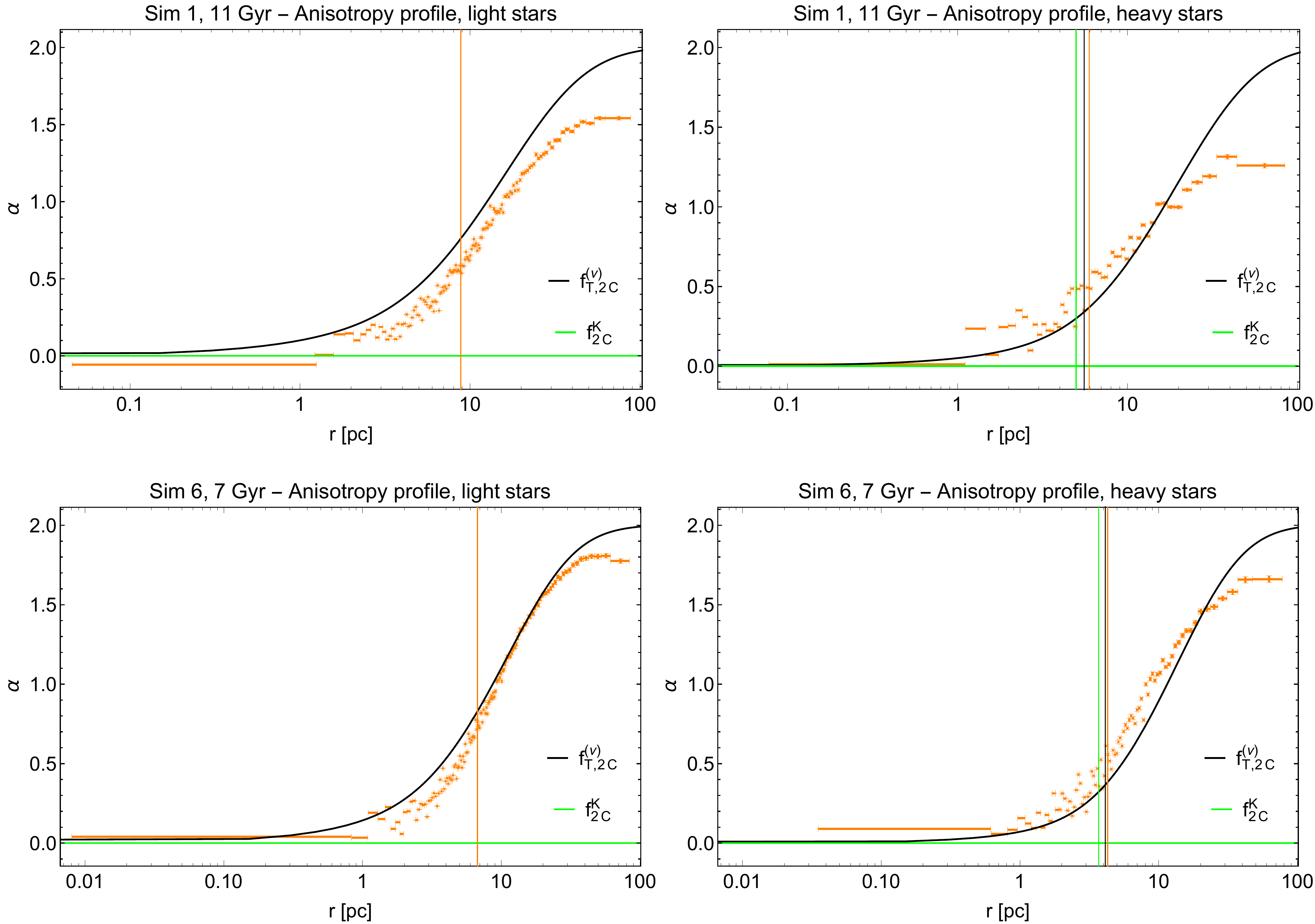}
							\caption{Anisotropy profiles for Sim 1 at 11 Gyr (upper panels) and  Sim 6 at 7 Gyr (lower panels) compared to those associated with best-fit two-component $f_T^{(\nu)}$ models (black) and the isotropic King models (green). Left panels: anisotropy profiles of light components. Right panels: anisotropy profiles of heavy components. The vertical lines represent the half-mass radius of the component under consideration in the simulated states and in the models. }\label{alpha_fv2c_2sys}
							
							\vspace*{\floatsep}
							\vspace*{\floatsep}
							\vspace*{\floatsep}
							\vspace*{\floatsep}
							
							\includegraphics[width=0.6\hsize]{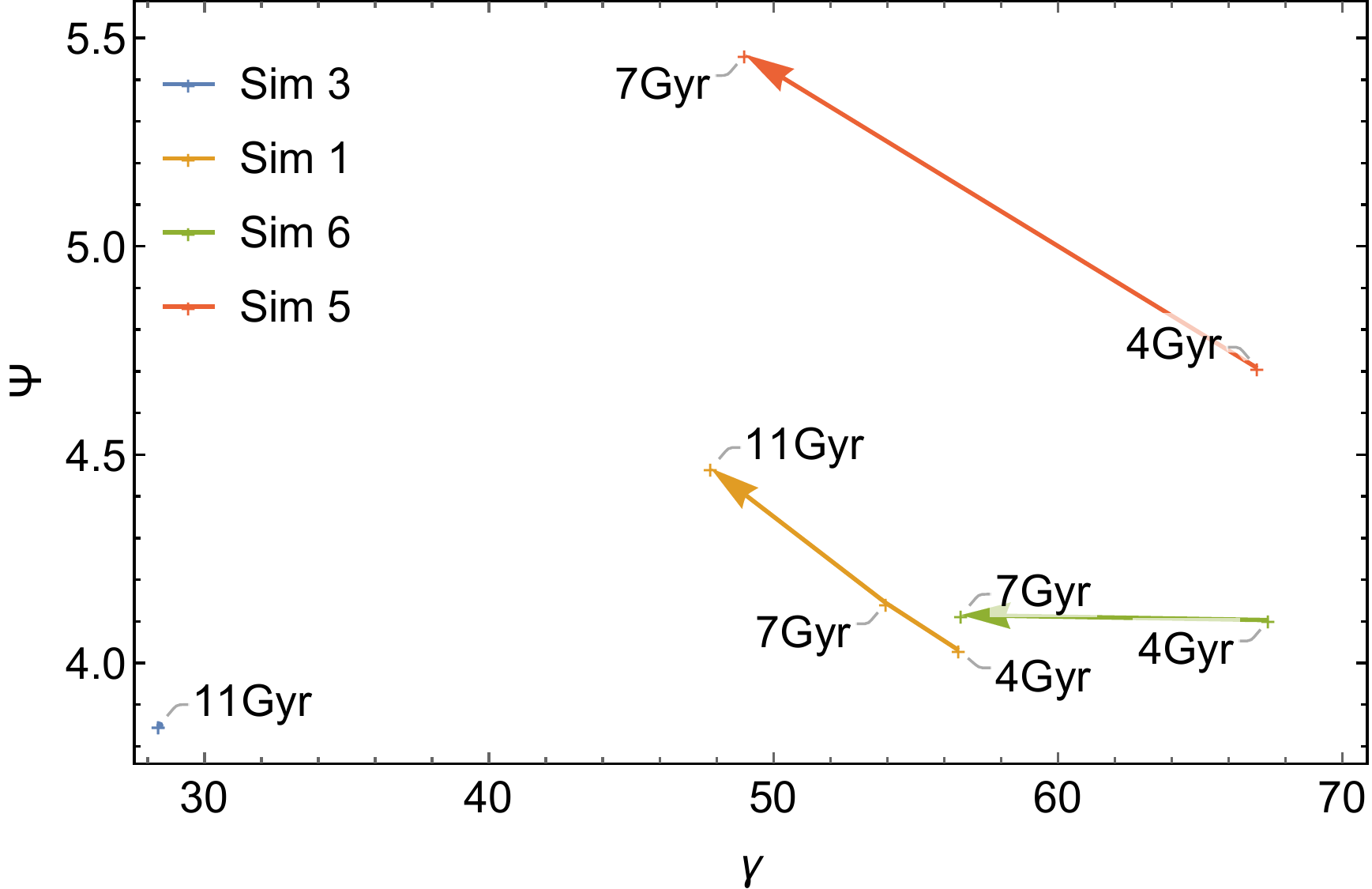}  
							\caption{Distribution of best-fit two-component $f_T^{(\nu)}$ models in the relevant parameter space. The arrows connect different snapshots of the same simulation.}\label{parameter_space}
							
						\end{figure*}
						
						For each of the two cases illustrated here in detail, we also compare the anisotropy profile of each component to that of the simulated state, as shown in Fig.~\ref{alpha_fv2c_2sys}. For the more relaxed system (Sim 6 at 7 Gyr), the best-fit $f^{(\nu)}_T$ model gives a good description of the local degree of anisotropy for both components, whereas for Sim 1 at 11 Gyr, the model anisotropy profile matches only that of the heavy component. These results suggest that the cumulative effects of collisions drive the systems toward a velocity distribution similar to that generated by violent relaxation.
						
						In App. \ref{App.A}, we summarize the results of the fits for one-component models: the $f^{(\nu)}_T$ models perform systematically better than the King models (as described in Tab. \ref{bestfit_1c}) but, in contrast
						to the two-component case, they do not reproduce the kinematic peak well at the center of the simulated states. As expected, the chi-squared values are higher than those of the two-component fit. Finally, in App. \ref{App.B}, we test the performance of two-component Michie-King models. We found that these models perform quite well, especially for the kinematical profiles,       but still worse than the $f^{(\nu)}_T$ models (as described in Tab. \ref{bestfit_2c_michie}). Given the structure of their distribution function, the Michie-King models give a reasonable description of the anisotropy profiles, although they exhibit a declining trend for $\alpha$ in the outer parts, which is not present in the simulated states.
						
						\subsubsection{Parameter space for two-component $f^{(\nu)}_T$ models}
						
						In Fig.~\ref{parameter_space}, we show the distribution of the best-fit two-component $f_T^{(\nu)}$ models in the relevant parameter space.
						The figure suggests that dynamical evolution drives the systems toward more truncated (lower values of $\gamma$) and more concentrated (higher values of $\Psi$) models. For Sim 1 and Sim 5, the evolution toward more concentrated models is more evident, whereas the simulated states relative to Sim 6 have about the same values of $\Psi$. This evolution in the parameter space might be related to a general trend in the direction of core collapse, which is usually attributed to the onset of the gravothermal catastrophe (\citealp{1968MNRAS.138..495L}).  A similar evolution trend in the parameter space (towards more truncated and more concentrated systems) was found for the LIMEPY models  (see \citealp{2016MNRAS.462..696Z} for one-component models and \citealp{2017MNRAS.470.2736P} for the multi-component case).  
						
						\section{Discussion and conclusions}
						
						In this paper, we consider a set of simulated states, meaning, selected snapshots taken from realistic Monte Carlo simulations (\citealp{2010MNRAS.407.1946D}) that incorporate both dynamical and stellar evolution. The simulated states were investigated by means of two-component $f_T^{(\nu)}$ (\citealp{2016A&A...590A..16D}) and \cite{1966AJ.....71...64K} models, in which the two components represent light (main sequence) stars and heavy stars (giants, remnants, and binaries), respectively. 
						The definition of the two-component models takes into account the presence of partial energy equipartition and mass segregation, which are expected to take place in many globular clusters as a result of collisions. The selected simulated states are characterized by different degrees of relaxation, with the relaxation parameter $n_{rel}=t_{age}/t_{rc}$ ranging from 2.5 to 64.9.
						A condition of only partial local equipartition is met at the center of the cluster, where it can be quantified by means of the parameter $\eta$ introduced by \cite{2013MNRAS.435.3272T}. All the simulated states exhibit $\eta \leq 0.27$, smaller than expected in the case of full central-energy equipartition ($\eta=0.5$). In turn{\tiny }, mass segregation is present also in the least relaxed systems, which indicates that this process sets in very efficiently. The complex interplay between energy equipartition and mass segregation has also been analyzed by \cite{2017MNRAS.464.1977W}; these authors quantify the mass segregation present in some simulations by measuring the gradient of the cluster's stellar-mass function, and refer to the parameter $\eta$ as a measure of energy equipartition. Here, in our simple two-component modeling, such complex interplay is summarized in Fig. \ref{slope_eta_relation}, which shows a linear relation between the slope of the half-mass radius profile, \textit{s}, and $\eta$. 
						
						By applying a combined density and kinematic chi-squared test to the two components, we find that two-component $f_T^{(\nu)}$ models provide a reasonable description of the density profiles of the simulated states. In particular, the two-component  $f_T^{(\nu)}$ models are able to reproduce the central peak in the velocity dispersion profiles (residuals are typically $<10\%$) and the increase in anisotropy profiles in more relaxed systems. This latter aspect suggests that the slow cumulative effects of relaxation processes lead the systems toward a velocity distribution that resembles that generated by collisionless violent relaxation.   By inspecting the evolution of simulated states in the parameter space, we may observe that dynamical evolution leads the systems toward more truncated and more concentrated models.
						In contrast, models based on the King distribution function do not offer a good representation of the simulated states. The density profiles of these models present a sharp truncation and high central densities, and the velocity dispersion profiles are not compatible with those measured in the simulations.
						
						We wish to emphasize that one limitation of the simulations considered in this paper is the lack of an exploration of the full effects of the tidal field. In this sense, the systems that are studied are quite isolated, which favors the onset of radial anisotropy in their outer regions. In the presence of a realistic tidal field, stars in radial orbits would be preferentially stripped and, as a result, the cluster would be more isotropic, at least in the outer parts, where, in the case of very strong tidal field, even tangential anisotropy could arise (e.g., see \citealp{2016MNRAS.462..696Z}, \citealp{2016MNRAS.461..402T}, \citealp{2017MNRAS.471.1181B}). Thus, we argue that the models developed here most likely represent an optimal tool for the clusters that underfill the volume associated with the boundary surface determined by the tidal interaction with the host galaxy.
						
						In conclusion, the two-component $f_T^{(\nu)}$ models appear to offer a realistic representation of a class globular clusters and a reasonably starting point to investigate dynamical mechanisms related to mass segregation and energy equipartition inside these stellar systems. The results thus set the basis for the application of the adopted family of models to diagnose the structural properties of nonrotating globular clusters, which will require a proper discussion of the relevant mass-to-light gradients and are given in a separate paper. 
						
						One future goal that is encouraged by the results of the present paper is to make an attempt to use the two-component $f_T^{(\nu)}$ models to interpret the real data of the new globular clusters diagnostics made possible by \textit{Gaia}. For the study of real systems, it is of primary importance to understand how mass segregation influences the local value of the mass-to-light ratio (e.g., see \citealp{2017MNRAS.469.4359B}), and thus the determination of the relevant parameters of the models considered. 
						From the theoretical point of view, a natural and interesting development of this work could be the construction of models able to take into account different degrees of anisotropy in the outer regions, following in detail the indications provided by the simulations. For this, a better understanding of the mechanisms leading to the anisotropy profiles resulting from collisionality would be desired.
						
						\section*{Acknowledgements}
						We would like to thank J.M.B. Downing for the Monte Carlo simulations used in this work. We wish to thank the Referee for the constructive report and C. Grillo, F. Pegoraro, and A.L. Varri for a number of useful discussions and suggestions. This work was partially supported by the Italian Miur, through a Prin contract. The Department of Physics of the University of Pisa is thanked for the kind hospitality extended to G. Bertin.

						\bibliographystyle{aa} 
						\bibliography{tbb19}

\begin{thebibliography}{56}
\expandafter\ifx\csname natexlab\endcsname\relax\def\natexlab#1{#1}\fi

\bibitem[{{Anderson} \& {King}(2003)}]{2003AJ....126..772A}
{Anderson}, J. \& {King}, I.~R. 2003, \aj, 126, 772

\bibitem[{{Bellazzini} {et~al.}(2012){Bellazzini}, {Bragaglia}, {Carretta},
  {Gratton}, {Lucatello}, {Catanzaro}, \& {Leone}}]{2012A&A...538A..18B}
{Bellazzini}, M., {Bragaglia}, A., {Carretta}, E., {et~al.} 2012, \aap, 538,
  A18

\bibitem[{{Bellini} {et~al.}(2014){Bellini}, {Anderson}, {van der Marel},
  {Watkins}, {King}, {Bianchini}, {Chanam{\'e}}, {Chandar}, {Cool}, {Ferraro},
  {Ford}, \& {Massari}}]{2014ApJ...797..115B}
{Bellini}, A., {Anderson}, J., {van der Marel}, R.~P., {et~al.} 2014, \apj,
  797, 115

\bibitem[{{Bellini} {et~al.}(2017){Bellini}, {Bianchini}, {Varri}, {Anderson},
  {Piotto}, {van der Marel}, {Vesperini}, \& {Watkins}}]{2017ApJ...844..167B}
{Bellini}, A., {Bianchini}, P., {Varri}, A.~L., {et~al.} 2017, \apj, 844, 167

\bibitem[{{Bertin} \& {Stiavelli}(1993)}]{1993RPPh...56..493B}
{Bertin}, G. \& {Stiavelli}, M. 1993, Reports on Progress in Physics, 56, 493

\bibitem[{{Bianchini} {et~al.}(2017{\natexlab{a}}){Bianchini}, {Sills}, \&
  {Miholics}}]{2017MNRAS.471.1181B}
{Bianchini}, P., {Sills}, A., \& {Miholics}, M. 2017{\natexlab{a}}, \mnras,
  471, 1181

\bibitem[{{Bianchini} {et~al.}(2017{\natexlab{b}}){Bianchini}, {Sills}, {van de
  Ven}, \& {Sippel}}]{2017MNRAS.469.4359B}
{Bianchini}, P., {Sills}, A., {van de Ven}, G., \& {Sippel}, A.~C.
  2017{\natexlab{b}}, \mnras, 469, 4359

\bibitem[{{Bianchini} {et~al.}(2016){Bianchini}, {van de Ven}, {Norris},
  {Schinnerer}, \& {Varri}}]{2016MNRAS.458.3644B}
{Bianchini}, P., {van de Ven}, G., {Norris}, M.~A., {Schinnerer}, E., \&
  {Varri}, A.~L. 2016, \mnras, 458, 3644

\bibitem[{{Bianchini} {et~al.}(2018){Bianchini}, {van der Marel}, {del Pino},
  {Watkins}, {Bellini}, {Fardal}, {Libralato}, \&
  {Sills}}]{2018MNRAS.481.2125B}
{Bianchini}, P., {van der Marel}, R.~P., {del Pino}, A., {et~al.} 2018, \mnras,
  481, 2125

\bibitem[{{Bianchini} {et~al.}(2013){Bianchini}, {Varri}, {Bertin}, \&
  {Zocchi}}]{2013ApJ...772...67B}
{Bianchini}, P., {Varri}, A.~L., {Bertin}, G., \& {Zocchi}, A. 2013, \apj, 772,
  67

\bibitem[{{Cordero} {et~al.}(2017){Cordero}, {H{\'e}nault-Brunet},
  {Pilachowski}, {Balbinot}, {Johnson}, \& {Varri}}]{2017MNRAS.465.3515C}
{Cordero}, M.~J., {H{\'e}nault-Brunet}, V., {Pilachowski}, C.~A., {et~al.}
  2017, \mnras, 465, 3515

\bibitem[{{Da Costa} \& {Freeman}(1976)}]{1976ApJ...206..128D}
{Da Costa}, G.~S. \& {Freeman}, K.~C. 1976, \apj, 206, 128

\bibitem[{{de Vita} {et~al.}(2016){de Vita}, {Bertin}, \&
  {Zocchi}}]{2016A&A...590A..16D}
{de Vita}, R., {Bertin}, G., \& {Zocchi}, A. 2016, \aap, 590, A16

\bibitem[{{Di Cecco} {et~al.}(2013){Di Cecco}, {Zocchi}, {Varri}, {Monelli},
  {Bertin}, {Bono}, {Stetson}, {Nonino}, {Buonanno}, {Ferraro}, {Iannicola},
  {Kunder}, \& {Walker}}]{2013AJ....145..103D}
{Di Cecco}, A., {Zocchi}, A., {Varri}, A.~L., {et~al.} 2013, \aj, 145, 103

\bibitem[{{Djorgovski}(1993)}]{1993ASPC...50..373D}
{Djorgovski}, S. 1993, in Astronomical Society of the Pacific Conference
  Series, Vol.~50, Structure and Dynamics of Globular Clusters, ed. S.~G.
  {Djorgovski} \& G.~{Meylan}, 373

\bibitem[{{Djorgovski} \& {Meylan}(1994)}]{1994AJ....108.1292D}
{Djorgovski}, S. \& {Meylan}, G. 1994, \aj, 108, 1292

\bibitem[{{Downing} {et~al.}(2010){Downing}, {Benacquista}, {Giersz}, \&
  {Spurzem}}]{2010MNRAS.407.1946D}
{Downing}, J.~M.~B., {Benacquista}, M.~J., {Giersz}, M., \& {Spurzem}, R. 2010,
  \mnras, 407, 1946

\bibitem[{Efron \& Tibshirani(1986)}]{efron1986}
Efron, B. \& Tibshirani, R. 1986, Statist. Sci., 1, 54

\bibitem[{{Gaia Collaboration} {et~al.}(2018){Gaia Collaboration}, {Helmi},
  {van Leeuwen}, {McMillan}, {Massari}, {Antoja}, {Robin}, {Lindegren},
  {Bastian}, {Arenou}, \& et~al.}]{2018A&A...616A..12G}
{Gaia Collaboration}, {Helmi}, A., {van Leeuwen}, F., {et~al.} 2018, \aap, 616,
  A12

\bibitem[{{Gieles} \& {Zocchi}(2015)}]{2015MNRAS.454..576G}
{Gieles}, M. \& {Zocchi}, A. 2015, \mnras, 454, 576

\bibitem[{{Giersz}(1998)}]{1998MNRAS.298.1239G}
{Giersz}, M. 1998, \mnras, 298, 1239

\bibitem[{{Giersz} {et~al.}(2013){Giersz}, {Heggie}, {Hurley}, \&
  {Hypki}}]{2013MNRAS.431.2184G}
{Giersz}, M., {Heggie}, D.~C., {Hurley}, J.~R., \& {Hypki}, A. 2013, \mnras,
  431, 2184

\bibitem[{{Goldsbury} {et~al.}(2013){Goldsbury}, {Heyl}, \&
  {Richer}}]{2013ApJ...778...57G}
{Goldsbury}, R., {Heyl}, J., \& {Richer}, H. 2013, \apj, 778, 57

\bibitem[{{Gunn} \& {Griffin}(1979)}]{1979AJ.....84..752G}
{Gunn}, J.~E. \& {Griffin}, R.~F. 1979, \aj, 84, 752

\bibitem[{{Harris}(1996)}]{1996yCat.7195....0H}
{Harris}, W.~E. 1996, VizieR Online Data Catalog, 7195

\bibitem[{{Harris}(2010)}]{2010arXiv1012.3224H}
{Harris}, W.~E. 2010, arXiv e-prints [\eprint[arXiv]{1012.3224}]

\bibitem[{{H{\'e}nault-Brunet} {et~al.}(2019){H{\'e}nault-Brunet}, {Gieles},
  {Sollima}, {Watkins}, {Zocchi}, {Claydon}, {Pancino}, \&
  {Baumgardt}}]{2019MNRAS.483.1400H}
{H{\'e}nault-Brunet}, V., {Gieles}, M., {Sollima}, A., {et~al.} 2019, \mnras,
  483, 1400

\bibitem[{{H{\'e}non}(1971)}]{1971Ap&SS..13..284H}
{H{\'e}non}, M. 1971, \apss, 13, 284

\bibitem[{{Jindal} {et~al.}(2019){Jindal}, {Webb}, \&
  {Bovy}}]{2019arXiv190311070J}
{Jindal}, A., {Webb}, J.~J., \& {Bovy}, J. 2019, arXiv e-prints
  [\eprint[arXiv]{1903.11070}]

\bibitem[{{Kacharov} {et~al.}(2014){Kacharov}, {Bianchini}, {Koch}, {Frank},
  {Martin}, {van de Ven}, {Puzia}, {McDonald}, {Johnson}, \&
  {Zijlstra}}]{2014A&A...567A..69K}
{Kacharov}, N., {Bianchini}, P., {Koch}, A., {et~al.} 2014, \aap, 567, A69

\bibitem[{{Kamann} {et~al.}(2018){Kamann}, {Husser}, {Dreizler}, {Emsellem},
  {Weilbacher}, {Martens}, {Bacon}, {den Brok}, {Giesers}, {Krajnovi{\'c}},
  {Roth}, {Wendt}, \& {Wisotzki}}]{2018MNRAS.473.5591K}
{Kamann}, S., {Husser}, T.~O., {Dreizler}, S., {et~al.} 2018, \mnras, 473, 5591

\bibitem[{{King}(1966)}]{1966AJ.....71...64K}
{King}, I.~R. 1966, \aj, 71, 64

\bibitem[{{Kroupa}(2001)}]{2001MNRAS.322..231K}
{Kroupa}, P. 2001, \mnras, 322, 231

\bibitem[{{Lardo} {et~al.}(2015){Lardo}, {Pancino}, {Bellazzini}, {Bragaglia},
  {Donati}, {Gilmore}, {Randich}, {Feltzing}, {Jeffries}, {Vallenari},
  {Alfaro}, {Allende Prieto}, {Flaccomio}, {Koposov}, {Recio-Blanco},
  {Bergemann}, {Carraro}, {Costado}, {Damiani}, {Hourihane}, {Jofr{\'e}}, {de
  Laverny}, {Marconi}, {Masseron}, {Morbidelli}, {Sacco}, \&
  {Worley}}]{2015A&A...573A.115L}
{Lardo}, C., {Pancino}, E., {Bellazzini}, M., {et~al.} 2015, \aap, 573, A115

\bibitem[{{Libralato} {et~al.}(2019){Libralato}, {Bellini}, {Piotto},
  {Nardiello}, {van der Marel}, {Anderson}, {Bedin}, \&
  {Vesperini}}]{2019ApJ...873..109L}
{Libralato}, M., {Bellini}, A., {Piotto}, G., {et~al.} 2019, \apj, 873, 109

\bibitem[{{Libralato} {et~al.}(2018){Libralato}, {Bellini}, {van der Marel},
  {Anderson}, {Watkins}, {Piotto}, {Ferraro}, {Nardiello}, \&
  {Vesperini}}]{2018ApJ...861...99L}
{Libralato}, M., {Bellini}, A., {van der Marel}, R.~P., {et~al.} 2018, \apj,
  861, 99

\bibitem[{{Lynden-Bell}(1967)}]{1967MNRAS.136..101L}
{Lynden-Bell}, D. 1967, \mnras, 136, 101

\bibitem[{{Lynden-Bell} \& {Wood}(1968)}]{1968MNRAS.138..495L}
{Lynden-Bell}, D. \& {Wood}, R. 1968, \mnras, 138, 495

\bibitem[{{McLaughlin} \& {van der Marel}(2005)}]{2005ApJS..161..304M}
{McLaughlin}, D.~E. \& {van der Marel}, R.~P. 2005, \apjs, 161, 304

\bibitem[{{Merritt}(1981)}]{1981AJ.....86..318M}
{Merritt}, D. 1981, \aj, 86, 318

\bibitem[{{Michie}(1963)}]{1963MNRAS.125..127M}
{Michie}, R.~W. 1963, \mnras, 125, 127

\bibitem[{{Miocchi}(2006)}]{2006MNRAS.366..227M}
{Miocchi}, P. 2006, \mnras, 366, 227

\bibitem[{{Peuten} {et~al.}(2017){Peuten}, {Zocchi}, {Gieles}, \&
  {H{\'e}nault-Brunet}}]{2017MNRAS.470.2736P}
{Peuten}, M., {Zocchi}, A., {Gieles}, M., \& {H{\'e}nault-Brunet}, V. 2017,
  \mnras, 470, 2736

\bibitem[{{Plummer}(1911)}]{1911MNRAS..71..460P}
{Plummer}, H.~C. 1911, \mnras, 71, 460

\bibitem[{{Sollima} {et~al.}(2019){Sollima}, {Baumgardt}, \&
  {Hilker}}]{2019MNRAS.485.1460S}
{Sollima}, A., {Baumgardt}, H., \& {Hilker}, M. 2019, \mnras, 485, 1460

\bibitem[{{Sollima} {et~al.}(2015){Sollima}, {Baumgardt}, {Zocchi}, {Balbinot},
  {Gieles}, {H{\'e}nault-Brunet}, \& {Varri}}]{2015MNRAS.451.2185S}
{Sollima}, A., {Baumgardt}, H., {Zocchi}, A., {et~al.} 2015, \mnras, 451, 2185

\bibitem[{{Spitzer}(1969)}]{1969ApJ...158L.139S}
{Spitzer}, Jr., L. 1969, \apjl, 158, L139

\bibitem[{{Tiongco} {et~al.}(2016){Tiongco}, {Vesperini}, \&
  {Varri}}]{2016MNRAS.461..402T}
{Tiongco}, M.~A., {Vesperini}, E., \& {Varri}, A.~L. 2016, \mnras, 461, 402

\bibitem[{{Trenti} \& {van der Marel}(2013)}]{2013MNRAS.435.3272T}
{Trenti}, M. \& {van der Marel}, R. 2013, \mnras, 435, 3272

\bibitem[{{van Albada}(1982)}]{1982MNRAS.201..939V}
{van Albada}, T.~S. 1982, \mnras, 201, 939

\bibitem[{{van der Marel} \& {Anderson}(2010)}]{2010ApJ...710.1063V}
{van der Marel}, R.~P. \& {Anderson}, J. 2010, \apj, 710, 1063

\bibitem[{{Vishniac}(1978)}]{1978ApJ...223..986V}
{Vishniac}, E.~T. 1978, \apj, 223, 986

\bibitem[{{Watkins} {et~al.}(2015){Watkins}, {van der Marel}, {Bellini}, \&
  {Anderson}}]{2015ApJ...803...29W}
{Watkins}, L.~L., {van der Marel}, R.~P., {Bellini}, A., \& {Anderson}, J.
  2015, \apj, 803, 29

\bibitem[{{Webb} \& {Vesperini}(2017)}]{2017MNRAS.464.1977W}
{Webb}, J.~J. \& {Vesperini}, E. 2017, \mnras, 464, 1977

\bibitem[{{Zocchi} {et~al.}(2012){Zocchi}, {Bertin}, \&
  {Varri}}]{2012A&A...539A..65Z}
{Zocchi}, A., {Bertin}, G., \& {Varri}, A.~L. 2012, \aap, 539, A65

\bibitem[{{Zocchi} {et~al.}(2016){Zocchi}, {Gieles}, {H{\'e}nault-Brunet}, \&
  {Varri}}]{2016MNRAS.462..696Z}
{Zocchi}, A., {Gieles}, M., {H{\'e}nault-Brunet}, V., \& {Varri}, A.~L. 2016,
  \mnras, 462, 696

\end{thebibliography}

						
						\clearpage
						\appendix
						
						
						\section{Fit by one-component models}
						\label{App.A}
						
						\begin{table*}[ht]
							\centering
							\caption{Best-fit parameters for one-component models}
							\resizebox{\textwidth}{!}{\begin{tabular}{ l c c c c c c c c c c }
									\hline \hline
									&       & \multicolumn{2}{c}{King models} & & \multicolumn{6}{c}{$f^{(\nu)}_T$ models}   \\  \cline{2-5} \cline{7-11}
									& $\Psi$ &  $\tilde{\chi}^2_{\rho_{tot}}$ & $\tilde{\chi}^2_{\sigma_{tot}}$ & $\tilde{\chi}^2_{tot}$ & &  $\Psi$  & $\gamma$  & $\tilde{\chi}^2_{\rho_{tot}}$ & $\tilde{\chi}^2_{\sigma_{tot}}$ & $\tilde{\chi}^2_{tot}$  \\ \hline 
									Sim 3, 11 Gyr & $5.850 \pm   0.005$ & 32.04 & 114.71 & 73.12 &   & $5.03 \pm 0.01$ & $26.0    \pm 0.2$ & 5.44 & 41.44 & 23.36 \\
									Sim 1, 4 Gyr & $ 6.638 \pm 0.004$  & 89.90  & 367.42  & 227.94 & & $4.95 \pm 0.02$ & $67.3 \pm 0.24$ & 9.93 & 52.59 & 31.16 \\
									Sim 1, 7 Gyr &$6.682 \pm 0.004$ & 139.75 & 350.46 & 244.33 & &$5.12 \pm 0.01$ & $61.8 \pm 0.3$ & 5.73 & 42.25 & 23.92\\
									Sim 1, 11 Gyr & $6.761 \pm  0.005$ & 81.80 & 242.02 & 161.38 & & $5.46 \pm 0.01$ & $55.6 \pm 0.3$ & 2.88 & 41.68 & 22.20  \\
									Sim 6, 4 Gyr & $7.151 \pm 0.004$ & 143.53 & 469.08 & 305.43 & & $5.26 \pm 0.02$ & $76.8 \pm 0.2$ & 3.00 & 49.93 & 26.39   \\
									Sim 5, 4 Gyr & $7.399 \pm 0.005$ & 143.38 & 496.77 & 318.96 &  & $5.92 \pm 0.01$ & $80.0 \pm      0.3$ &  5.15 &  56.26 &       30.60   \\
									Sim 6, 7 Gyr & $7.507 \pm 0.003$ & 159.22 & 374.67 & 266.15 & & $6.20 \pm 0.01$ & $58.0    \pm 0.2$ & 6.93 & 49.93 & 28.35  \\
									Sim 5, 7 Gyr & $7.717 \pm 0.006$ & 127.29 & 502.05 & 313.54 & & $7.06 \pm 0.01$ & $52.4 \pm 0.2$ & 10.83 & 55.91 & 33.25   \\
									\hline
								\end{tabular}}%
								\label{bestfit_1c}%
							\end{table*}%
							
							The simplest and most common way to apply stellar dynamics to the study of stellar systems is to fit the available data by means of one-component self-consistent models. When such a procedure is adopted, it is assumed that the stellar populations may include stars of different masses, but that they are homogeneous. In the introduction, we noted that this picture may be viable for collisionless systems in the absence of dark matter, but is inherently inappropriate when collisionality induces effects of mass segregation and equipartition. Below, we illustrate the performance of one-component dynamical models on data taken from the simulated states that were studied in the main text by means of two-component models. In particular, we compare the fits obtained from King models to those from  $f^{(\nu)}_T$ models.
							
\begin{figure*}[ht]
								\centering
								\includegraphics[width=1\textwidth]{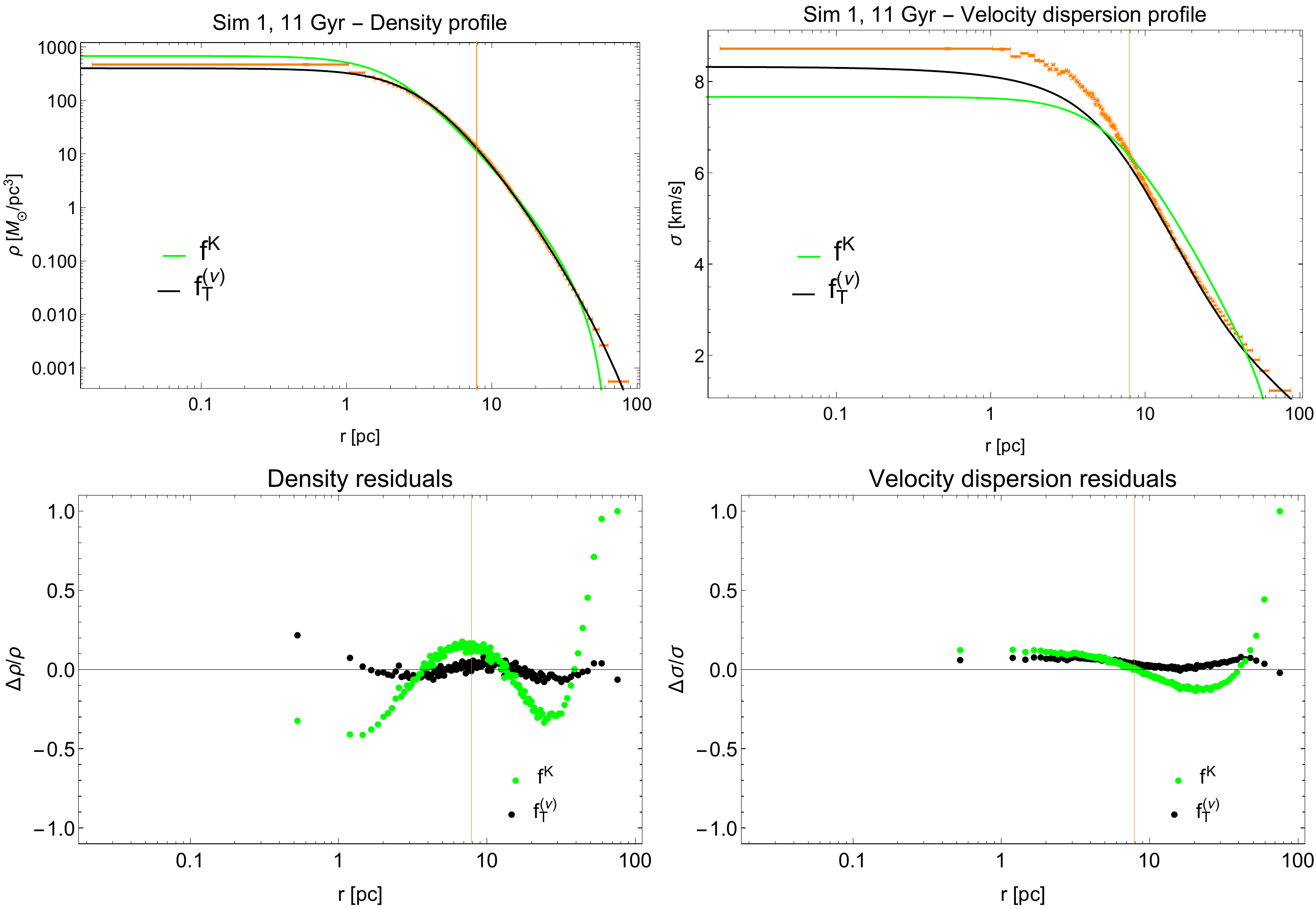}  
								\caption{ Best-fit profiles and residuals for Sim 1 at 11 Gyr for one-component King models (green) and for one-component $f^{(\nu)}_T$ models (black). Upper panels: density (left) and velocity dispersion (right) profiles. Lower panels: density (left) and velocity dispersion (right) residuals.}\label{fig_1c_sim1_11}
							\end{figure*}

\begin{figure*}[ht]
								\centering
								\includegraphics[width=1\textwidth]{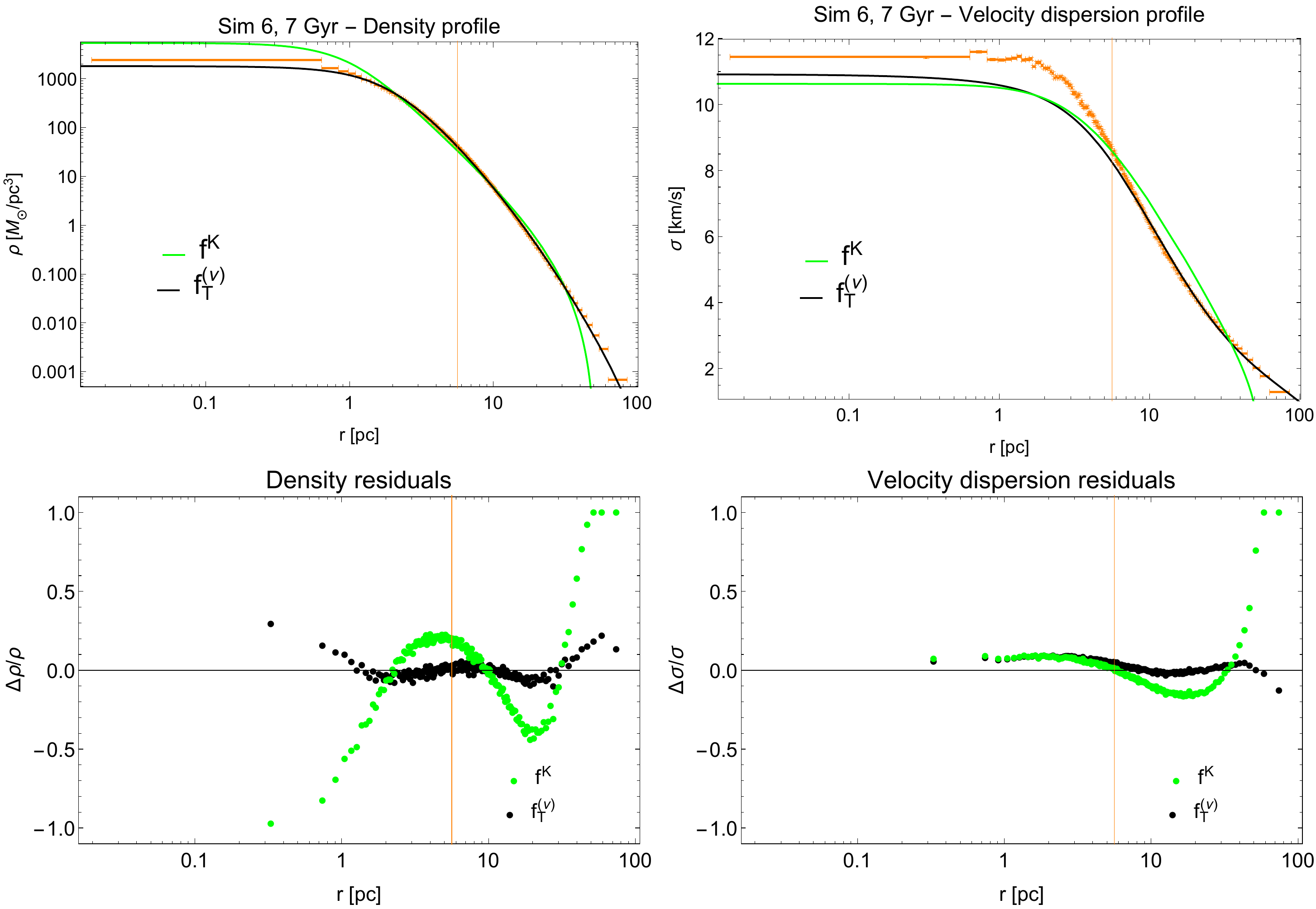}  
								\caption{ Best-fit profiles and residuals for Sim 6 at 7 Gyr for one-component King models (green) and for one-component $f^{(\nu)}_T$ models (black). Upper panels: density (left) and velocity dispersion (right) profiles. Lower panels: density (left) and velocity dispersion (right) residuals.}\label{fig_1c_sim6_7}
							\end{figure*}

							In Fig.~\ref{fig_1c_sim1_11} and Fig~ \ref{fig_1c_sim6_7}, we show the best-fit density and velocity dispersion profiles for Sim 1 at 11 Gyr and for Sim 6 at 7 Gyr, respectively. The best-fit values of the dimensionless parameters (and their formal errors) and the reduced chi-squared for one-component models are listed in Tab.~\ref{bestfit_1c}. The set of simulated states is listed in order of increasing relaxation.  A trend of increasing $\Psi$ for more relaxed systems is apparent for both dynamical models. 
							For any given simulated state, the values of $\Psi$ and $\gamma$ associated with the one-component models are systematically larger than those reported in Tab. \ref{bestfit_2c} for the corresponding fits by two-component models.
							
\begin{figure*}[ht]
								\centering
								\includegraphics[width=1\textwidth]{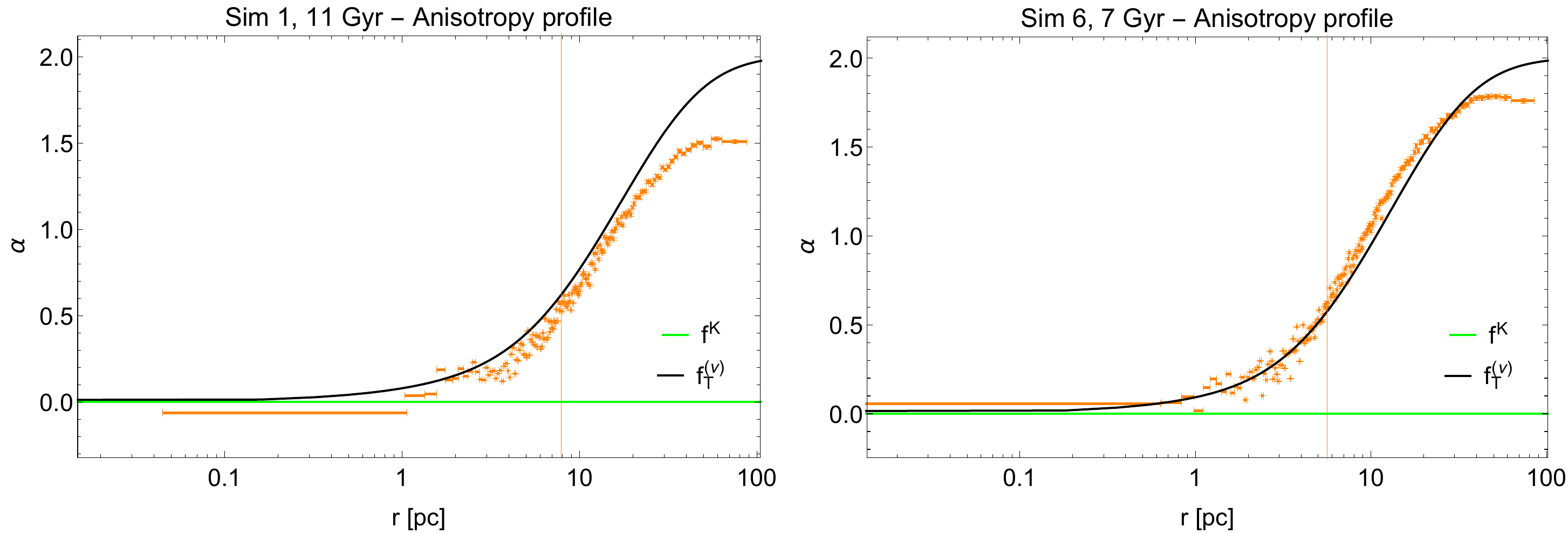}  
								\caption{Anisotropy profiles of Sim 1 at 11 Gyr (left panel) and  Sim 6 at 7 Gyr (right panel) compared to those associated with best-fit one-component $f_T^{(\nu)}$ models (black) and the isotropic King models (green).}\label{alpha_fv1c_2sys}
							\end{figure*}

							For each best-fit model illustrated here, we also compare the anisotropy profile to that of the simulated state, as shown in Fig.~\ref{alpha_fv1c_2sys}. Interestingly, the $f^{(\nu)}_T$ models give a good description of the pressure anisotropy for relaxed systems, such as Sim 6 at 7 Gyr, whereas Sim 1 at 11 Gyr presents a lower degree of radial anisotropy in the outer regions. (This confirms the view that collisions drive the system toward a velocity distribution that resembles that generated by the completely different mechanism of violent relaxation.)
							
							Apparently, one-component King models do not perform well, although they are found to perform slightly better for less relaxed cases. They are able to fit the density profile reasonably well, but the models exhibit an exceedingly sharp density truncation. The one-component King models cannot reproduce the central peak or the general qualitative behavior of the velocity dispersion profile. This is probably related to the fact that King models have isotropic velocity distributions, whereas the simulated states are characterized by significant (radially-biased) pressure anisotropy. 
							
							The one-component $f^{(\nu)}_T$ models show a significant improvement with respect to the King models. Indeed, the $f_T^{(\nu)}$ density profiles provide a better representation of the simulated states, except for the central shell, with a welcome mild density truncation in the outer parts. As for the velocity dispersion profiles, the models do not reproduce the central peak that characterizes the simulations, but their overall shape matches the properties of the simulated states better than the best-fit King models. 
							
							The values of $\tilde{\chi}_{tot}^2$ are very high, in particular for the kinematic fit of the King models. This is partly due to the low values of the formal uncertainties on the data points, but it certainly marks a failure of the one-component models in representing the structure of the simulated states. We conclude that the results of the fits by one-component models confirms that the $f^{(\nu)}_T$ models provide a step in the right direction, in relation to the structural properties of realistic simulations of globular clusters. However, the significant discrepancies that are noted simply underline the need to incorporate effects of mass segregation and energy equipartition that only multi-component models may be able to handle.

							\section{A comparison with two-component Michie-King Models}
							\label{App.B}
							For the two simulated states of our sample {(so chosen in order to be representative of different relaxation conditions), Sim 1 at 11 Gyr ($n_{rel}=8.3$), and Sim 6 at 7 Gyr ($n_{rel}=23.6$), we also compared the performance of the two-component $f^{(\nu)}_T$ models to that of the Michie-King (\citeyear{1963MNRAS.125..127M}) two-component models, defined by the following distribution function:
								
								\begin{equation} \label{fmichie}
								f^{MK}_i(E,J) = \begin{cases} A_{i} \exp{\left(-a_i \frac{J^2}{2 r^2_{a}}\right)} \big\{\exp{\left[-a_i (E-E_t)\right]}-1 \big\} \quad \rm{for} \quad E < E_t 
								\\ 0 \quad \rm{for} \quad E \geq  E_t \end{cases},
								\end{equation}
								
								\noindent where $A_i$, $a_i$, are positive constants referring to the i-th component and $r_a$ is the anisotropy radius, which is the radius at which the anisotropy function $\alpha$, defined in Eq. (\ref{alpha_def}), equals unity.  Following a procedure analogous to that outlined in Sect. \ref{sec_2cmodels} for the two-component $f^{(\nu)}_T$ models, we reduced the set of parameters identifying this family of models to two dimensionless free parameters, which is the concentration of the light component, $\Psi=-a_1(\Phi(0)-E_t)$, and $\gamma_M=a^{1/2} /(4 \pi G A_1 r_a^2)$. In this case, assumption (3) of Sect. \ref{sec_2cmodels} is replaced by the choice that the two components are characterized by the same anisotropy radius.
								
								We performed a combined chi-squared analysis of the density and velocity dispersion profiles, following the same procedure as in Sect. \ref{sec_fitting_2c}. In contrast to the other models considered in this paper, in this case, the length scale is set by equating the anisotropy radius of the light component, which appears explicitly in the models, to that of the simulated states. 
								
								In Tab.~\ref{bestfit_2c_michie}, we report the best-fit values of the dimensionless parameters and the reduced chi-squared for the comparison between the two-component models and the simulated states considered, together with those obtained for the two-component $f^{(\nu)}_T$ models. The Michie-King  models show values of $\tilde{\chi}^2$ significantly higher than those of the $f^{(\nu)}_T$ models. 
								
								For Sim 6 at 7 Gyr, we also represent the best-fit associated profiles in Fig. \ref{fig_2c_sim6_7_density_m} and Fig. \ref{fig_2c_sim6_7_velocity_m}, compared to those of the best-fit $f^{(\nu)}_T$ models.
								The two-component Michie-King models appear to perform better than the two-component King models (discussed in the main text), because their phase space structure has the qualitatively appealing feature of an anisotropy profile of the right kind; this special feature is further optimized by setting, in the fitting procedure, the model anisotropy radius of the light component to be equal to that of the simulated data. Indeed, this positive aspect is illustrated by the excellent performance in the kinematical fit of Fig. \ref{fig_2c_sim6_7_velocity_m}. However, the performance in fitting the density profiles (Fig. \ref{fig_2c_sim6_7_density_m}) is not as satisfactory, which explains the high values of chi-squared that are found.
								
								\begin{table*}[h]
									\centering
									\caption{Best-fit parameters for two-component Michie-King and $f^{(\nu)}_T$ models.}
									\resizebox{\textwidth}{!}{
										\begin{tabular}{l c c c c c c c c c c c c c c c}
											\hline \hline
											&       & \multicolumn{4}{c}{Michie-King models} & & & \multicolumn{8}{c}{$f^{(\nu)}_T$ models}   \\  \cline{2-8} \cline{10-16}
											& $\Psi$ & $\gamma_M$ & $\tilde{\chi}^2_{\rho_1}$ &  $\tilde{\chi}^2_{\sigma_1}$ &  $\tilde{\chi}^2_{\rho_2}$ & $\tilde{\chi}^2_{\sigma_2}$ &  $\tilde{\chi}^2_{tot}$ & &$\Psi$ &  $\gamma$ &  $\tilde{\chi}^2_{\rho_1}$  &  $\tilde{\chi}^2_{\sigma_1}$ & $\tilde{\chi}^2_{\rho_2}$ & $\tilde{\chi}^2_{\sigma_2}$ &  $\tilde{\chi}^2_{tot}$  \\ \hline
											
											Sim 1, 11 Gyr & $3.400 \pm 0.006$ & $27.9 \pm 0.8$ & 55.16 & 33.12  & 15.59 & 63.01 & 42.09 & & $4.467 \pm 0.002$ & $47.8 \pm 0.2$& 13.58  & 18.05 & 7.73 & 15.89 & 14.46  \\
											
											Sim 6, 7 Gyr & $3.732 \pm 0.005$ & $40.9 \pm 0.2$ & 53.79 & 16.35 & 87.74 & 18.10 & 41.17 &  & $4.115 \pm 0.003$ & $56.6 \pm 0.2$ & 3.68 & 21.35  & 7.05 & 6.59 & 10.23  \\
											
											\hline
										\end{tabular}}%
										\label{bestfit_2c_michie}%
									\end{table*}%

									We also tested the performance of the Michie-King models in describing the anisotropy profile of each component, as shown in Fig.~\ref{alpha_fv2c_2sys_m}. Despite the absence of the outer decrease of the anisotropy profile in the simulated state, the best-fit Michie-King model performs reasonably well in describing the local degree of anisotropy, especially for the light component.

									\begin{figure*}
										\centering
										\includegraphics[width=1\textwidth]{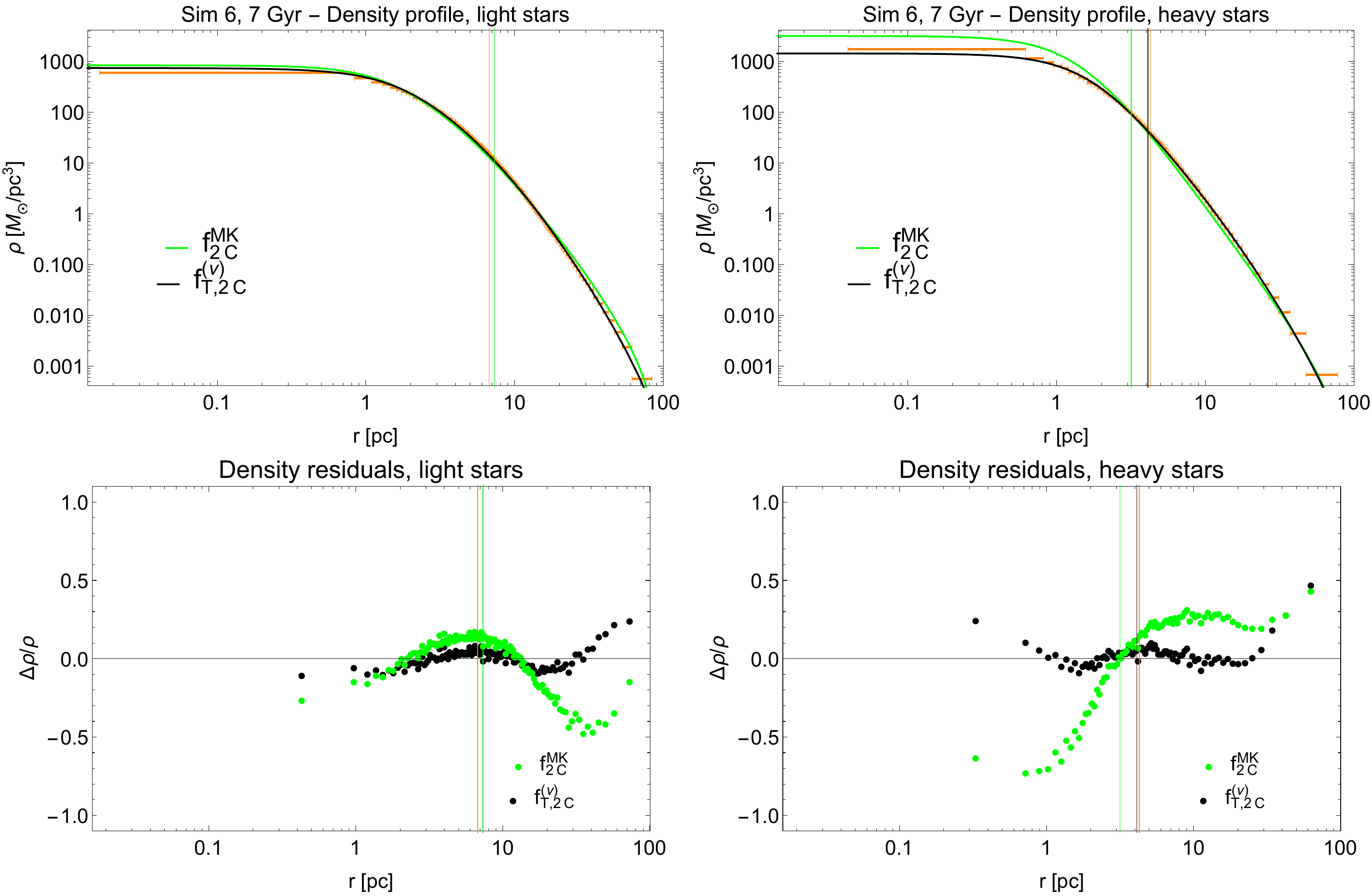}  
										\caption{Best-fit profiles and residuals for Sim 6 at 7 Gyr for two-component Michie-King models (green) and for two-component $f^{(\nu)}_T$ models (black). Upper panels: density profile for light component (left) and for heavy component (right). Lower panels: density residuals for light component (left) and for heavy component (right). The vertical lines represent the half-mass radius of the component under consideration in the simulated state and in the models.}\label{fig_2c_sim6_7_density_m}
									\end{figure*}
									\begin{figure*}[!htb]
										\centering
										\includegraphics[width=\hsize]{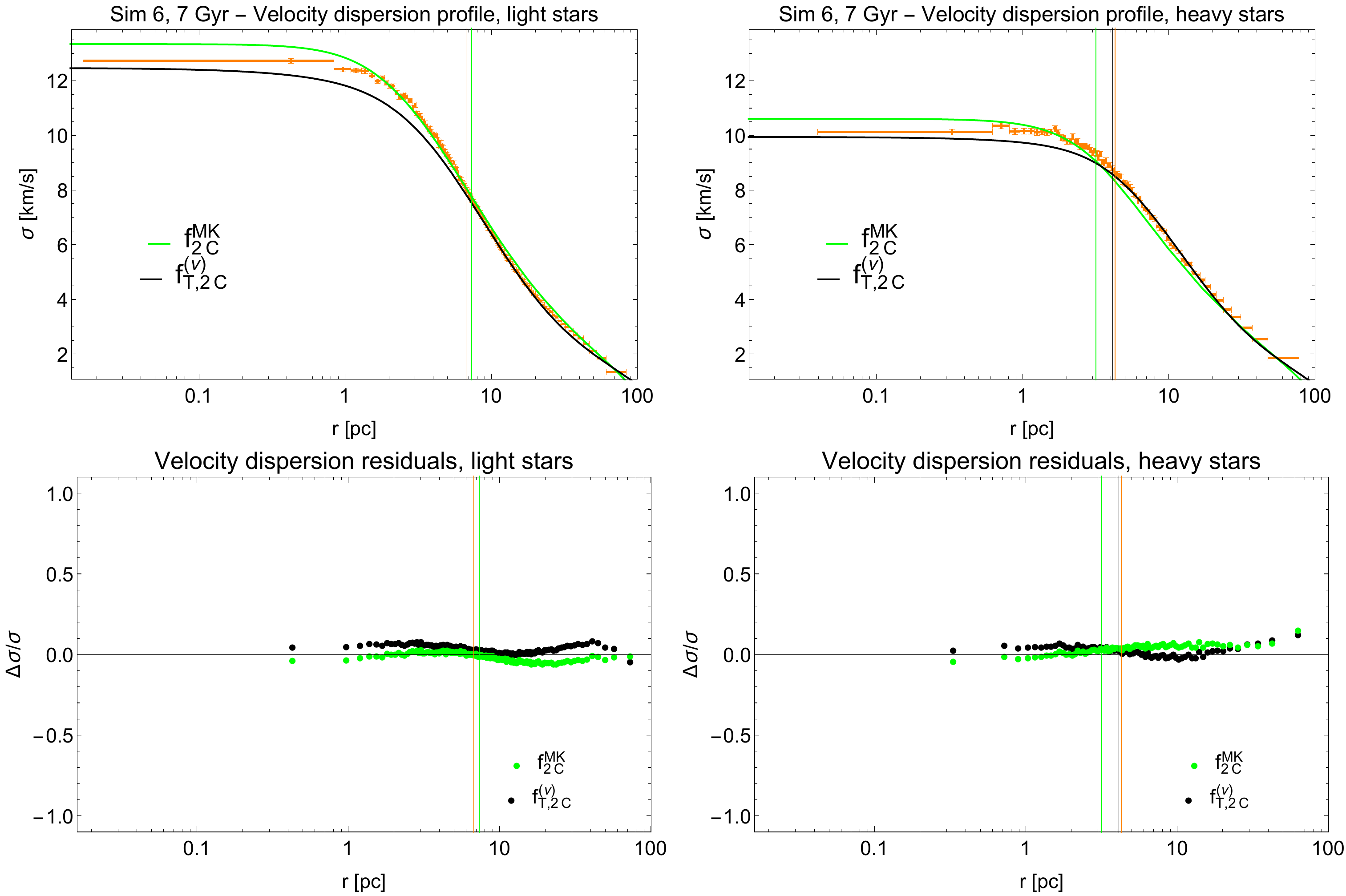}  
										\caption{Best-fit profiles and residuals for Sim 6 at 7 Gyr for two-component Michie-King models (green) and for two-component $f^{(\nu)}_T$ models (black). Upper panels: velocity dispersion profile for light component (left) and for heavy component (right). Lower panels: velocity dispersion residuals for light component (left) and for heavy component (right). The vertical lines represent the half-mass radius of the component under consideration in the simulated state and in the models. }\label{fig_2c_sim6_7_velocity_m}
									\end{figure*}
									
									\begin{figure*}[!htb]
										\centering
										\includegraphics[width=1\textwidth]{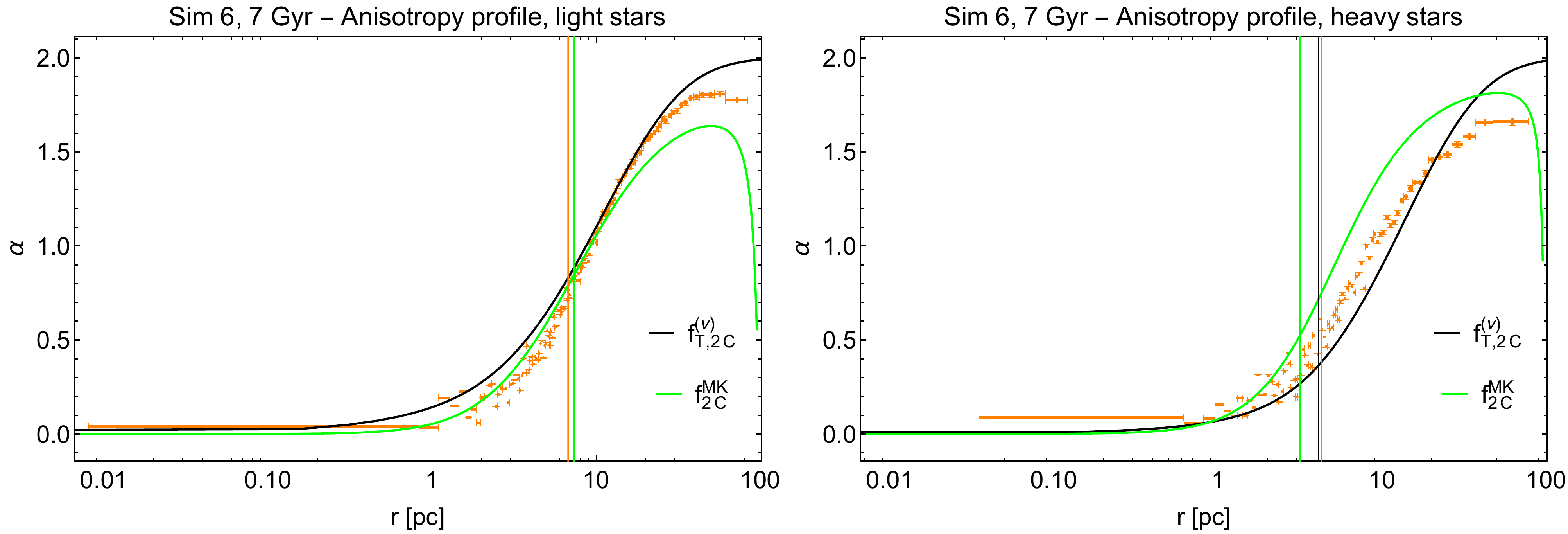} 
										\caption{ Anisotropy profiles Sim 6 at 7 Gyr compared to those associated with best-fit two-component $f_T^{(\nu)}$ models (black) and Michie-King models (green). Left panel: anisotropy profile of light component. Right panel: anisotropy profiles of heavy component. The vertical lines represent the half-mass radius of the component under consideration in the simulated states and in the models. }\label{alpha_fv2c_2sys_m}
									\end{figure*}
									
								\end{document}